\documentclass[12pt,a4paper]{article}
\pdfoutput=1

\usepackage{ifthen} 
\newboolean{pdflatex}
\setboolean{pdflatex}{true} 

\newboolean{articletitles}
\setboolean{articletitles}{true} 

\newboolean{uprightparticles}
\setboolean{uprightparticles}{false} 

\newboolean{inbibliography}
\setboolean{inbibliography}{false} 

\def\paperauthors{LHCb collaboration} 
\def\papertitle{Central exclusive production of \jpsi and \psitwos mesons in $pp$ collisions at $\sqrt{s}=13\tev$} 
\def\papercopyright{\the\year\ CERN for the benefit of the LHCb collaboration} 
\def\paperlicence{CC-BY-4.0 licence}
\def\paperlicenceurl{https://creativecommons.org/licenses/by/4.0/}

\input{preamble}

\begin{document}

\renewcommand{\thefootnote}{\fnsymbol{footnote}}
\setcounter{footnote}{1}


\begin{titlepage}
\pagenumbering{roman}

\vspace*{-1.5cm}
\centerline{\large EUROPEAN ORGANIZATION FOR NUCLEAR RESEARCH (CERN)}
\vspace*{1.5cm}
\noindent
\begin{tabular*}{\linewidth}{lc@{\extracolsep{\fill}}r@{\extracolsep{0pt}}}
\ifthenelse{\boolean{pdflatex}}
{\vspace*{-3.1cm}\mbox{\!\!\!\includegraphics[width=.14\textwidth]{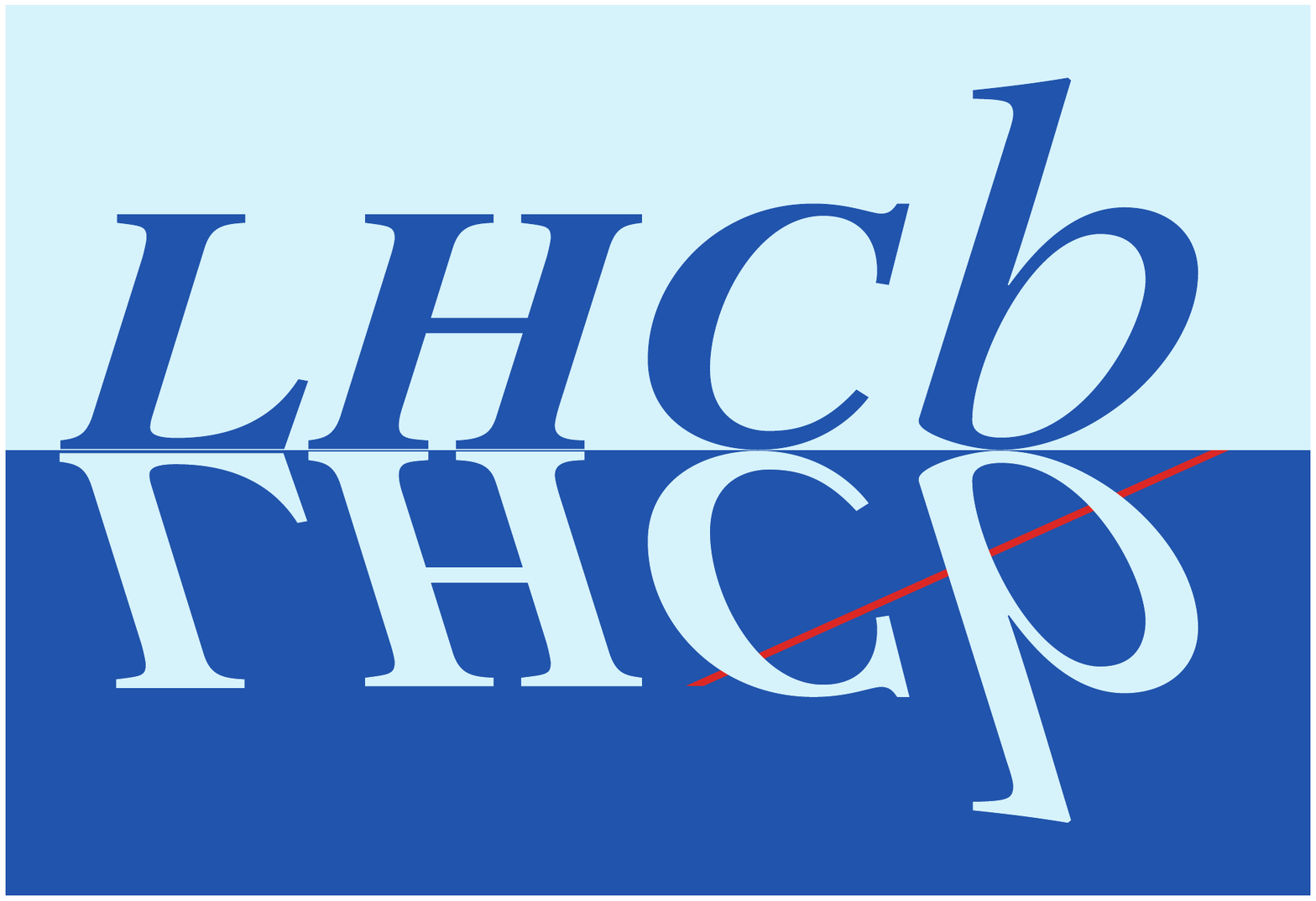}} & &}%
{\vspace*{-1.2cm}\mbox{\!\!\!\includegraphics[width=.12\textwidth]{lhcb-logo.eps}} & &}%
\\
 & & CERN-EP-2018-152 \\  
 & & LHCb-PAPER-2018-011 \\  
 & & \today \\ 
 & & \\
\end{tabular*}

\vspace*{2cm}

{\normalfont\bfseries\boldmath\huge
\begin{center}
  \papertitle 
\end{center}
}

\vspace*{1cm}

\begin{center}
\paperauthors\footnote{Authors are listed at the end of this paper.}
\end{center}

\vspace{\fill}

\begin{abstract}
  \noindent
  Measurements are reported of the central exclusive production of \jpsi and \psitwos
mesons in $pp$ collisions
at a centre-of-mass energy of 13\tev.
Backgrounds are significantly reduced compared to previous measurements 
made at lower energies
through the use
of new forward shower counters.
The products of the cross-sections and the branching fractions for the decays to dimuons, 
where both muons are within the pseudorapidity range
$2.0<\eta<4.5$, are measured to be
$$
 \begin{array}{rcl}
   \sigma_{J/\psi\rightarrow\mu^+\mu^-}&=&435 \pm 18 \pm 11 \pm 17 {\rm \ pb}\\ 
   \sigma_{\psi(2S)\rightarrow\mu^+\mu^-}&=&11.1 \pm 1.1 \pm 0.3 \pm 0.4 {\rm \ pb}.\\ 
\end{array}
 $$ 
The first uncertainties are statistical, the second are
systematic, and the third are due to the luminosity determination.
The cross-sections are also measured differentially for meson rapidities between 2.0 and 4.5.
Good agreement is observed with theoretical predictions.
Photoproduction cross-sections are
derived and compared to
previous experiments, and a deviation from a pure power-law extrapolation of lower energy data is observed.
\end{abstract}

\vspace*{1.0cm}

\begin{center}
  Published in JHEP 
\end{center}

\vspace{\fill}

{\footnotesize 
\centerline{\copyright~\papercopyright. \href{\paperlicenceurl}{\paperlicence}.}}
\vspace*{2mm}

\end{titlepage}


\newpage
\setcounter{page}{2}
\mbox{~}
%
%
%
%

\cleardoublepage


\renewcommand{\thefootnote}{\arabic{footnote}}
\setcounter{footnote}{0}




\pagestyle{plain} 
\setcounter{page}{1}
\pagenumbering{arabic}

%

\section{Introduction}

Central exclusive production (CEP)~\cite{review} of a vector meson 
in $pp$ collisions is a diffractive process in which the protons 
remain intact and the meson is produced through the fusion of 
a photon and a colourless strongly coupled object, the so-called pomeron.
For charmonia production, the cross-section can be predicted in 
perturbative quantum chromodynamics (QCD) and at the leading order (LO)
is proportional to the square of the gluon parton distribution function (PDF),
which ensures a steep rise in the photoproduction cross-section with the centre-of-mass energy of the photon-proton system, $W$.
Therefore, measurements of CEP of the \jpsi and \psitwos mesons provide not only a test of perturbative QCD but also 
 probe the pomeron, and constrain the gluon PDF.

Elastic photoproduction of charmonia has been measured in
fixed target experiments~\cite{e401,e516,e687}, in electron-proton~\cite{h1_latest,h1_jpsi,zeus_jpsi,h1psi}, $p\bar{p}$~\cite{cdf},
and proton-lead collisions~\cite{alice}.
The LHCb collaboration has previously 
measured the CEP of the \jpsi and \psitwos mesons in $pp$ collisions at a centre-of-mass energy $\sqrt{s}=7\tev$~\cite{lhcb7}.
In this paper, those results are extended to $\sqrt{s}=13\tev$ and
charmonia are measured up to $W=2\tev$, 
the highest energy yet explored.
This corresponds to 
probing the gluon PDF down to a fractional momentum of the proton, described by the Bjorken variable
$x\approx2\times10^{-6}$, a scale at which saturation effects may
become visible \cite{Goncalves:2016sqy}.

For diffractive processes, 
the dependence of the cross-section on the four-momentum transfer squared, $t$, is exponential with
a slope $b$ related to the 
transverse size of the interaction region.
In Regge theory~\cite{Collins:1977jy,Donnachie:1994zb}, 
$b$ varies with $W$ according to 
$b=b_0+4\alpha^\prime\log(W/W_0)$, where $b_0$ is the slope measured at an energy $W_0$.
Measurements at HERA determined $\alpha^\prime=0.164\pm0.041\gev^{-2}$
with $b_0=4.63^{+0.07}_{-0.17} \gev^{-2}$ for \jpsi photoproduction at $W_0=90\gev$~\cite{h1_jpsi}.
In $pp$ collisions at $\sqrt{s}=7\tev$, LHCb measured $b=5.70\pm 0.11\gev^{-2}$ at an average value of $W=750\gev$~\cite{lhcb7}. According to Regge theory, a value of $b\approx6.1\gev^{-2}$ is expected
for \jpsi production in $pp$ collisions at $\sqrt{s}=13\tev$.
In inelastic \jpsi production when proton dissociation occurs, the fall-off with $t$
is more gradual. In contrast, 
the nonresonant ultraperipheral electromagnetic CEP of dimuons, 
produced through photon-photon fusion, 
peaks strongly at low $t$ values.
Therefore, the $t$ dependence of the cross-section can be
 used to distinguish 
and study different production mechanisms. 

 This paper presents measurements of the cross-section for central exclusive
production of charmonia with rapidity, $y$, between 2.0 and 4.5, 
and follows the methodology of the LHCb analysis at $\sqrt{s}=7\tev$~\cite{lhcb7}.
Exclusive charmonium candidates are selected 
through their characteristic signature at 
a hadron collider: a $pp$ interaction devoid of any activity
save the charmonium that is reconstructed from its decay to two muons.
The addition of new forward shower counters (\herschel) \cite{LHCb-DP-2016-003} extends the pseudorapidity
region in which particles can be vetoed and roughly 
halves the number of background events 
compared to the previous measurement.

The LHCb detector is outlined in Sec.~\ref{sec:det}
while the data and selection criteria are described in Sec.~\ref{sec:sel}. 
The cross-section calculation is detailed in Sec.~\ref{sec:cs} and
  systematic uncertainties are presented in Sec.~\ref{sec:sys}. The cross-section results
for $pp\rightarrow p\jpsi p$ and $pp\rightarrow p\psitwos p$ processes
and derived photoproduction cross-sections for 
$\gamma p\rightarrow \jpsi p$ and $\gamma p\rightarrow \psitwos p$
are presented in Sec.~\ref{sec:results}.
Conclusions are given in Sec.~\ref{sec:conclude}.

\section{Detector, data samples and triggers}
\label{sec:det}

The \lhcb detector~\cite{Alves:2008zz,LHCb-DP-2014-002} is a single-arm forward
spectrometer covering the \mbox{pseudorapidity} range $2<\eta <5$,
designed for the study of particles containing \bquark or \cquark
quarks. The detector includes a high-precision tracking system
consisting of a silicon-strip vertex detector (VELO) surrounding the $pp$
interaction region, a large-area silicon-strip detector located
upstream of a dipole magnet with a bending power of about
$4{\mathrm{\,Tm}}$, and three stations of silicon-strip detectors and straw
drift tubes placed downstream of the magnet.
The tracking system provides a measurement of momentum, \ptot, of charged particles with
a relative uncertainty that varies from 0.5\% at low momentum to 1.0\% 
at 200\gev.\footnote{Natural units with $c=1$ are used throughout.}
Photons, electrons and hadrons are identified by a calorimeter system consisting of
scintillating-pad (SPD) and preshower detectors, an electromagnetic
calorimeter and a hadronic calorimeter. 
Muons are identified by a
system composed of alternating layers of iron and multiwire
proportional chambers~\cite{LHCb-DP-2012-002}.

The pseudorapidity coverage is extended by forward shower counters
consisting of 
five planes of scintillators  
 with three planes at 114, 19.7 and 7.5~m upstream
of the interaction point, and two downstream at 20 and 114~m.
At each location there are four quadrants of scintillators, 
whose information is recorded in
every beam crossing by photomultiplier tubes, 
giving a total of 20 channels in \herschel.
These are calibrated using data taken without beams circulating
at the end of each LHC fill.
The pseudorapidity ranges covered by VELO and \herschel are different. For VELO, the region is $-3.5<\eta<-1.5$ and $2<\eta<5$,
and for \herschel, the region is $-10<\eta<-5$ and $5<\eta<10$.

A data set corresponding to an integrated luminosity of $204\pm 8$\invpb
in $pp$ collisions at $\sqrt{s}=13\tev$ is used in this analysis.
The average number of $pp$ interactions per beam crossing, $\mu$, 
is 1.1, thus in about half of visible interactions
there is only a single $pp$ collision and the CEP process is uncontaminated by pile-up.
The online event selection is performed by a trigger
that consists of two different stages. First, there is a hardware stage, 
which requires less than 30 deposits in the SPD and at least one muon 
with a transverse momentum, \pt, above 200\mev. It is
followed by a software stage, which applies a full event
reconstruction and requires fewer than ten 
reconstructed tracks, at least one of which is identified as a muon.

Simulated signal events are generated using SuperCHIC v2.02~\cite{superchic},
where the \jpsi and \psitwos mesons are transversely polarised.
The \jpsi meson can also originate from 
exclusive \chic decays, which are also
generated with SuperCHIC, or from \psitwos decays, which are handled by PYTHIA~\cite{pythia}.
The LPAIR generator~\cite{lpair} is used to generate
dimuons produced through the electromagnetic photon-photon fusion process.
The
interaction of the generated particles with the detector, and the detector response,
are implemented using the \geant
toolkit~\cite{Allison:2006ve, *Agostinelli:2002hh} as described in
Ref.~\cite{LHCb-PROC-2011-006}.

\section{Event selection}
\label{sec:sel}
The selection of candidate signal events 
is similar to that used in the previous LHCb analysis~\cite{lhcb7}.
Two reconstructed muons are required in the region $2.0<\eta<4.5$, 
with an invariant mass 
within $\pm 65 \mev$ of the known 
\jpsi or \psitwos mass~\cite{PDG2017} and $p^2_{\rm T}$ of the  
reconstructed meson below 0.8$\gev^2$.
The mass and $p^2_{\rm T}$ requirements are both chosen to reject background while ensuring good signal efficiency, the evaluations of which are described in Sec.~\ref{sec:pur} and \ref{sec:eff}.

Events with additional VELO tracks or photons with transverse energies above 200\mev are vetoed.
Events with significant deposits in \herschel are removed.
The \herschel response is described using a variable $\chi_{\rm HRC}^2$ that quantifies the activity
above noise, taking account of correlations between the counters.

\begin{figure}
  \begin{center}
    \includegraphics[width=10cm]{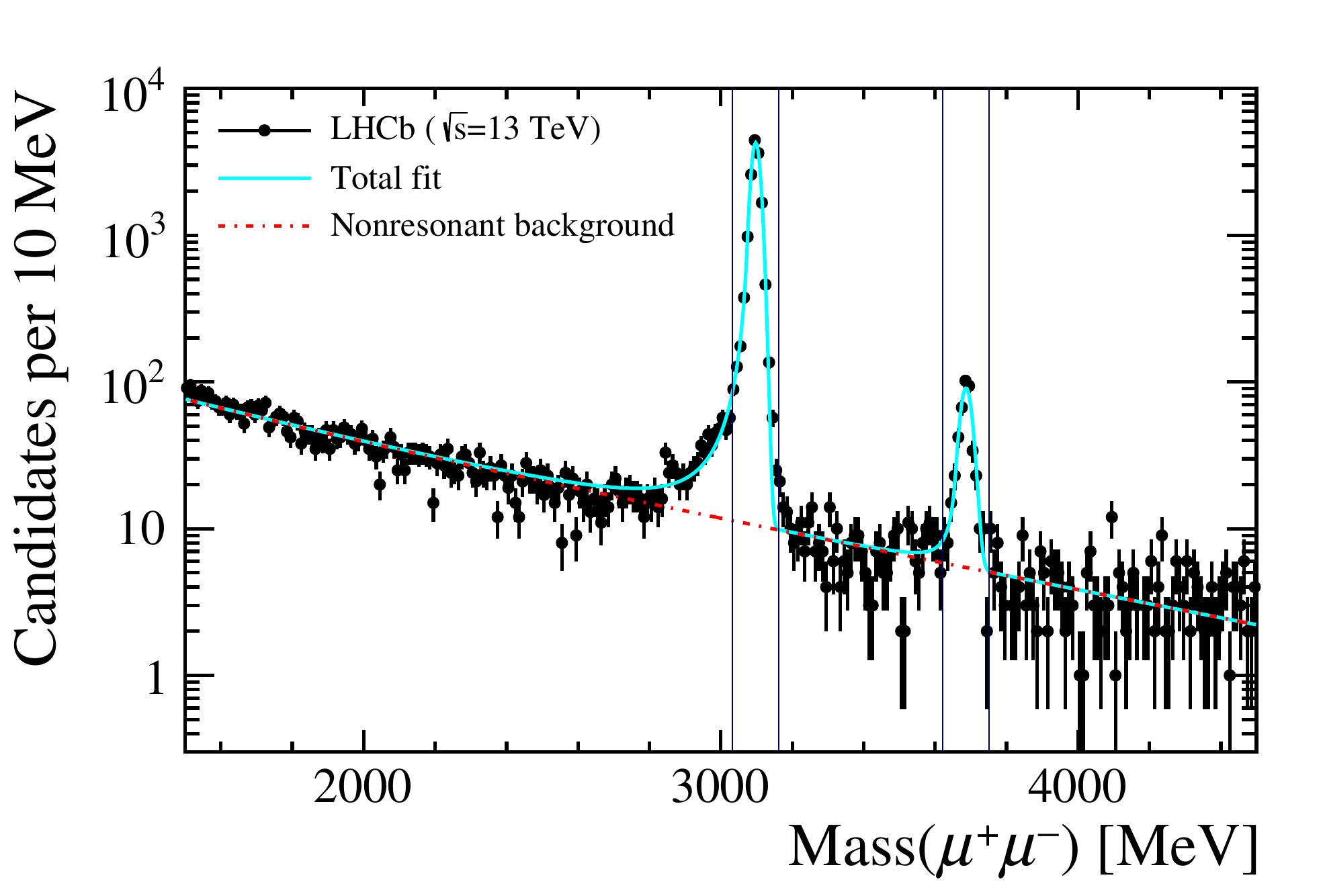}
\end{center}
  \caption{Invariant mass distribution of dimuon candidates.
The \jpsi and \psitwos
mass windows of the signal regions are indicated by the vertical lines.}
\label{fig:mass}
\end{figure}

\begin{figure}[!t]
  \begin{center}
    \includegraphics[width=10cm]{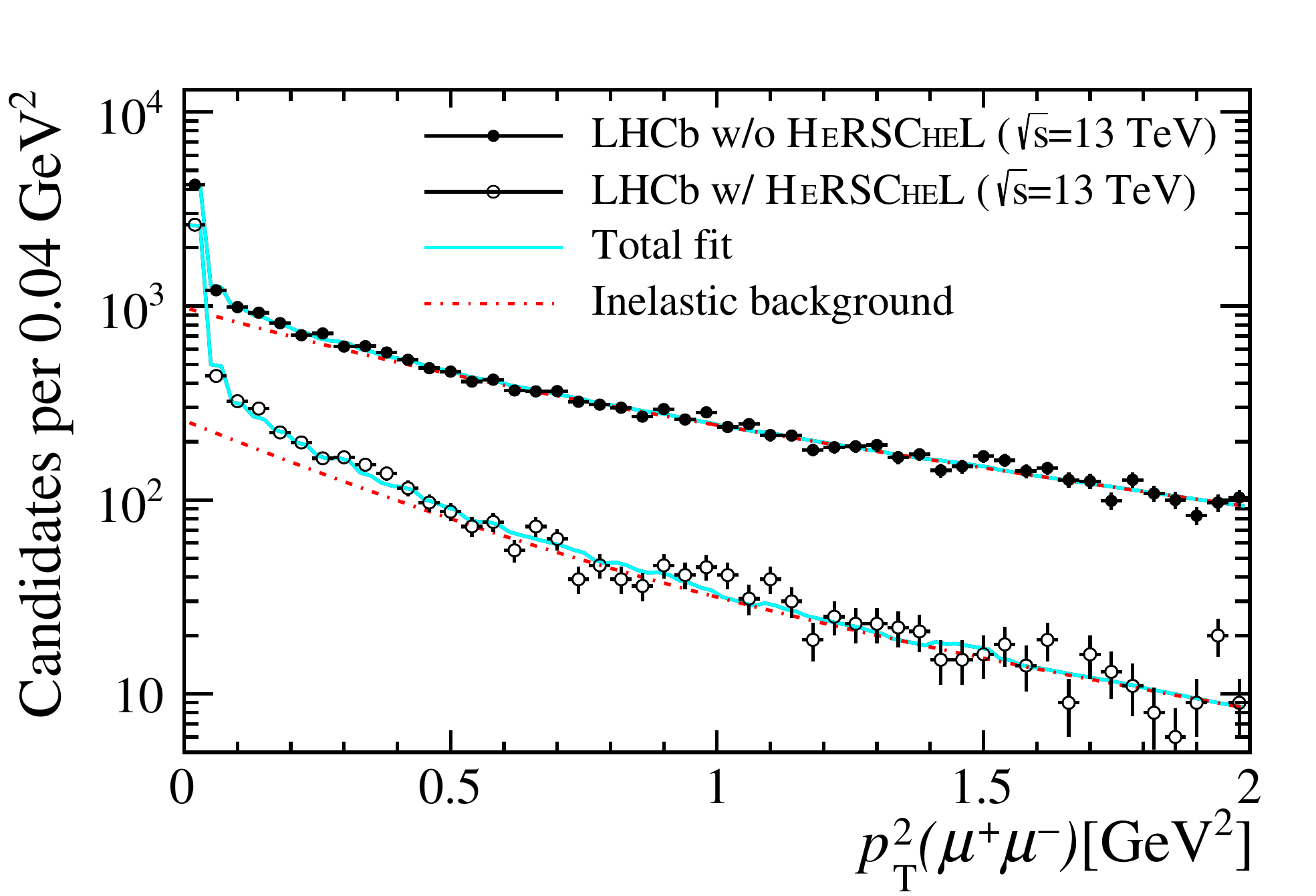}
\end{center}
\caption{Transverse  momentum squared for dimuons in the nonresonant region.
The upper distributions are without any requirement on \herschel: the lower are with the
\herschel veto applied. The total fit includes the electromagnetic CEP signal events as described by the LPAIR generator as well as the inelastic background.
}
\label{fig:qedpt2}
\end{figure}

The invariant mass, $M$, of all candidates 
without the mass-window requirement applied
is shown in Fig.~\ref{fig:mass}.
The data in the nonresonance regions (when $1500 <M<2700\mev$, $3200 <M<3500\mev$ and $3800 <M<8000\mev$) are 
candidates for electromagnetic 
CEP dimuons produced by photon-photon fusion 
and constitute an important calibration sample.
The $p_{\rm T}^2$ distribution of these dimuons with and without the requirement on $\chi_{\rm HRC}^2$ is shown in
Fig.~\ref{fig:qedpt2} and is significantly peaked towards low values due to the
long-range electromagnetic interaction.
The fraction of electromagnetic CEP events in this
sample is determined from a fit to the $p_{\rm T}^2$ distribution with two components: 
a signal shape taken from simulated events and an inelastic background modelled with
the sum of two exponential functions.

\begin{figure}[!t]
\begin{center}
\includegraphics[width=10cm]{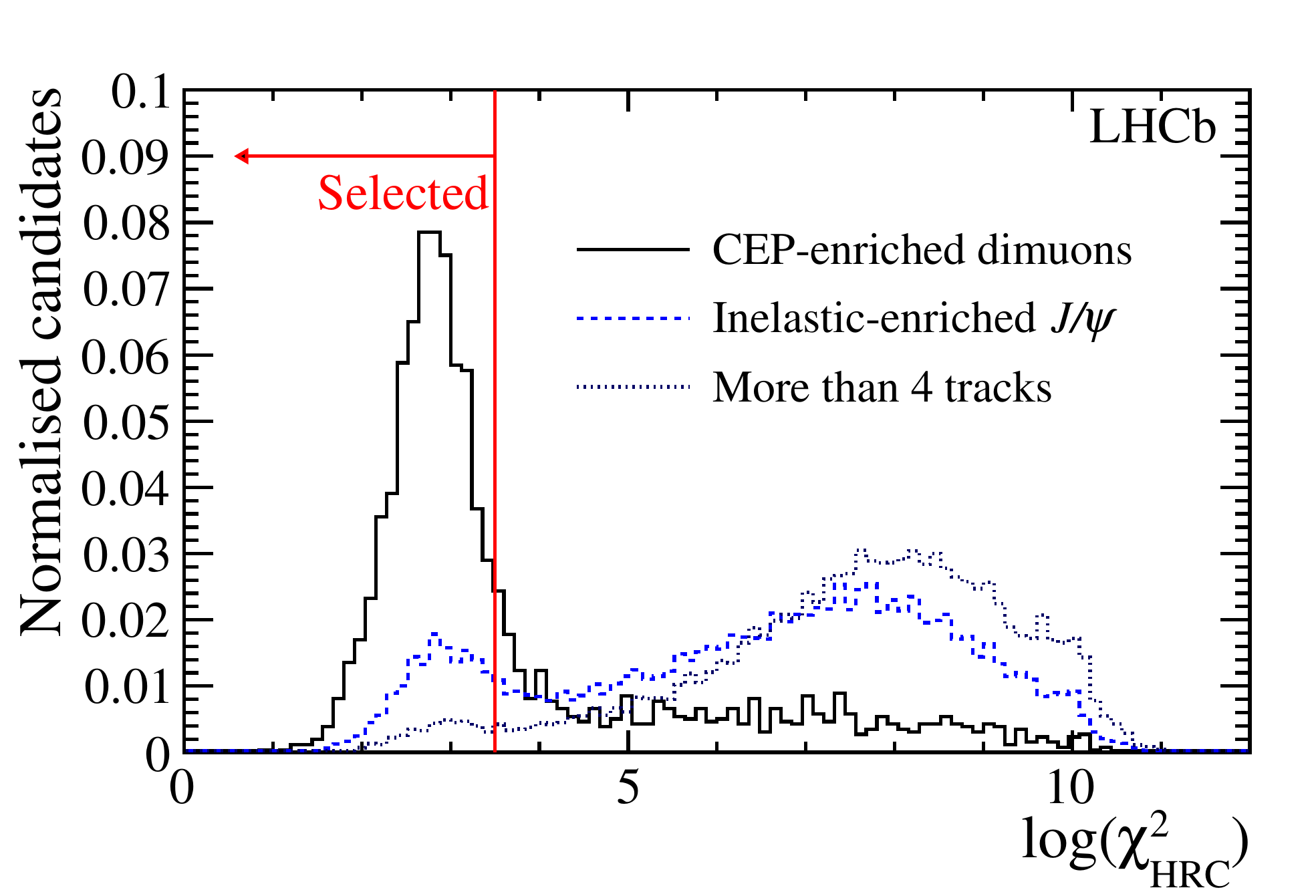}
\end{center}
\caption{Distributions, normalised to unit area,
of the logarithm of the discriminating variable $\chi_{\rm HRC}^2$
that is related to activity in \herschel. The response to three classes of
events, as described in the text, is shown.  
The selection requirement for the analysis is indicated by the red vertical line and the arrow.} \label{fig:her_cep}
\end{figure}

The power of \herschel to discriminate CEP events 
can be seen in Fig.~\ref{fig:her_cep}, which shows
the distributions of $\chi_{\rm HRC}^2$ for three classes of low-multiplicity-triggered events.
The first class is CEP-enriched dimuons: events in
the nonresonant dimuon sample with $p^2_{\rm T}<0.01\gev^2$, which has a purity of 97\% for electromagnetic CEP events.
The second class, inelastic-enriched \jpsi, applies the nominal \jpsi selections but requires $p^2_{\rm T}>1\gev^2$,
thus selecting inelastic events with proton dissociation.
The third class consists of events with more than four tracks reconstructed.
Figure~\ref{fig:her_cep} shows that CEP-enriched events have lower values of
$\chi_{\rm HRC}^2$.  
To select exclusive \jpsi and \psitwos candidates, it is required that
$\log(\chi_{\rm HRC}^2)<3.5$; this value is chosen in order to
 minimise the combined statistical and systematic uncertainty on the total cross-sections.
After the event selections, there are 14\,753 \jpsi signal candidates and 440 \psitwos signal candidates remaining.

\pagebreak

The estimation of the signal efficiency, $\epsilon_{\rm H}$, for the requirement 
$\log(\chi_{\rm HRC}^2)<3.5$ is described in Sec.~\ref{sec:hrceff}.  Using this, 
Sec.~\ref{sec:pur} explains how the purity of the signal sample is estimated.
The signal efficiency of all selection requirements is detailed in Sec.~\ref{sec:eff}.

\subsection{\herschel efficiency of selecting signal events}
\label{sec:hrceff}

The efficiency for the veto on \herschel activity is estimated from data
using the nonresonant calibration sample.
The fits to the $p^2_{\rm T}$ distributions in Fig.~\ref{fig:qedpt2} 
give the numbers of electromagnetic CEP events 
 with and without the \herschel veto.  The ratio of these gives the
 efficiency of the veto, which is determined to be $\epsilon_{\rm H}=0.723\pm0.008$. The signal loss includes in particular a contribution from events where there is an additional primary interaction only seen in the \herschel detector, as well as spill-over from previous collisions, electronic noise and calibration effects, as discussed in Ref.~\cite{LHCb-DP-2016-003}.
This efficiency, measured
using the nonresonant sample, is applicable to any CEP process, with the same veto, collected in this data-taking period.
\subsection{Purity of signal sample}
\label{sec:pur}
Three background sources are considered: nonresonant dimuon production; 
feed-down of CEP $\chi_{cJ}(1P)$ or \psitwos to \jpsi mesons and other undetected particles; and nonexclusive events where the proton dissociates but the remnants remain undetected.  

The amount of nonresonant background is determined from the fit shown in Fig.~\ref{fig:mass},
where the signals are modelled with two Crystal Ball functions~\cite{CBall} 
and the nonresonant background with the sum of two exponential functions.
This background is estimated to contribute a fraction of
$0.009\pm0.001$ to the \jpsi and $0.161\pm0.018$ to the \psitwos samples.

The \psitwos feed-down background in the \jpsi selection is determined using simulated events 
that have been normalised to have the same yield as the $\psi(2S)\rightarrow\mumu$ signal in data
and is estimated to contribute a fraction $0.015\pm0.001$ to the \jpsi samples.
The $\chi_{cJ}(1P)$ feed-down background
is determined using a data calibration sample,
 which contains events that pass the nominal \jpsi selection, 
except instead of zero photons,
it is required that there is exactly one reconstructed photon with a transverse energy above 200\mev.
The numbers of $\chi_{c0}(1P),~\chi_{c1}(1P)$, and $\chi_{c2}(1P)$ candidates 
in this calibration sample are determined from a fit to the invariant mass
of the dimuon plus photon system.
These are scaled
by the ratio of \jpsi to $J/\psi+\gamma$ candidates in the corresponding simulated $\chi_{cJ}(1P)$ sample 
 from which it is esimated that
a fraction of $0.005\pm0.001$ of the \jpsi candidate sample 
is due to feed-down from $\chi_{c0}(1P)$ mesons, 
$0.002\pm0.001$ from $\chi_{c1}(1P)$ mesons,
and $0.038\pm0.002$ from $\chi_{c2}(1P)$ mesons.
The total feed-down ratio from \psitwos and $\chi_{cJ}(1P)$ mesons is $0.060\pm0.002$,
to be compared to $0.101\pm0.009$ in the previous analysis~\cite{lhcb7}:
the addition of \herschel
suppresses events with proton dissociation, which are more numerous in 
the double-pomeron-exchange process that mediates $\chi_{cJ}(1P)$ production.

The fraction of nonexclusive events due to proton dissociation is determined through
the $p_{\rm T}^2$ distribution of the \jpsi and the \psitwos candidates, after a background subtraction
to remove contributions coming from the electromagnetic nonresonant and feed-down backgrounds. 
The electromagnetic component is shown in Fig.~\ref{fig:qedpt2}, while the feed-down shape is taken from the $J/\psi+\gamma$ calibration sample.
The background-subtracted $p_{\rm T}^2$ distribution consists of two remaining components:  signal and proton dissociation background.
Since $t\approx -p_{\rm T}^2$, approximately exponential distributions with different slopes
 are expected for each.  In the previous analysis~\cite{lhcb7}, each was modelled
by an exponential function whose
slope was a free parameter.
The presence of the \herschel detector however now allows these shapes to be determined
from data, thus reducing the model dependence of the result.

\begin{figure}[!t]
\begin{center}
\includegraphics[width=7.5cm]{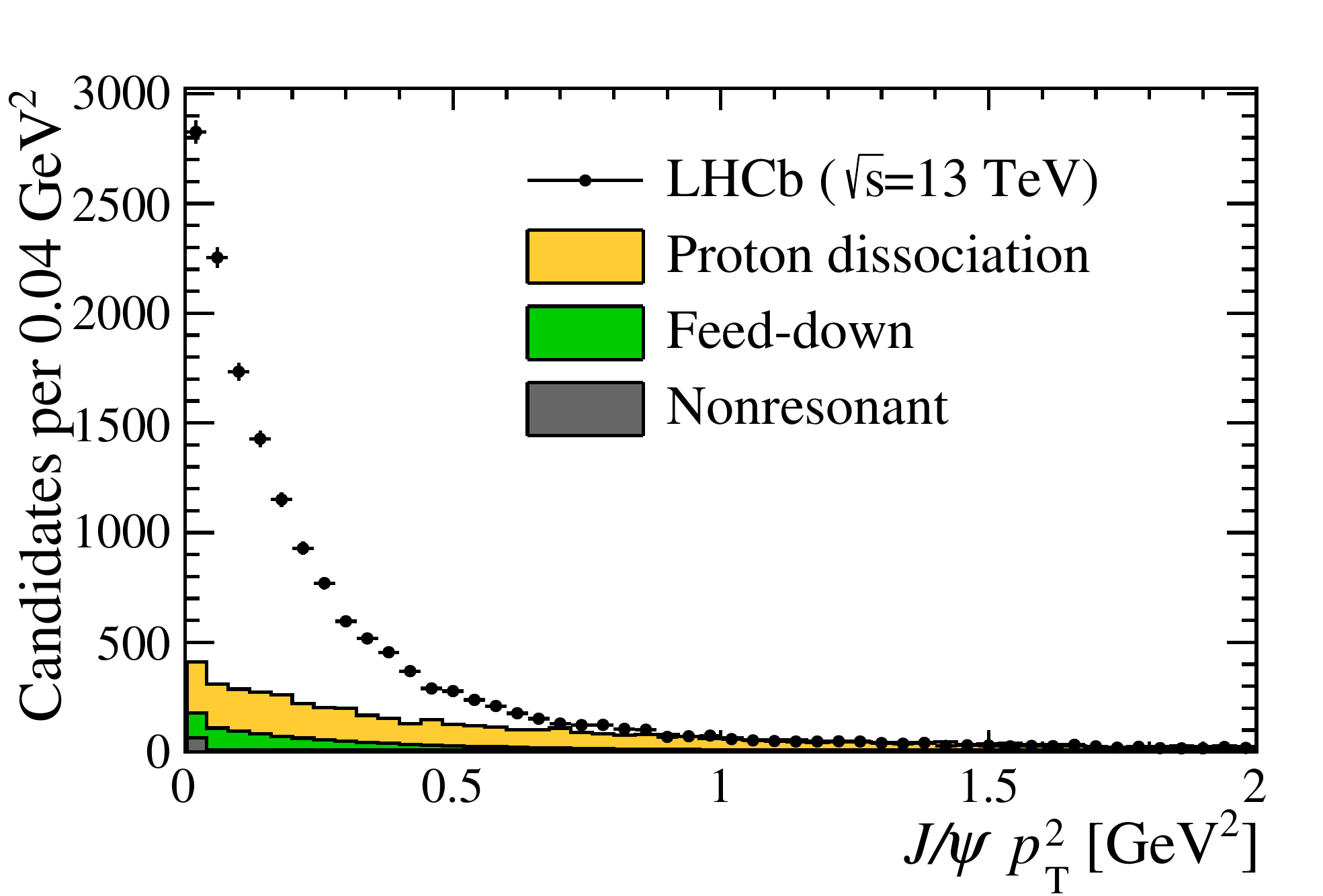}
\includegraphics[width=7.5cm]{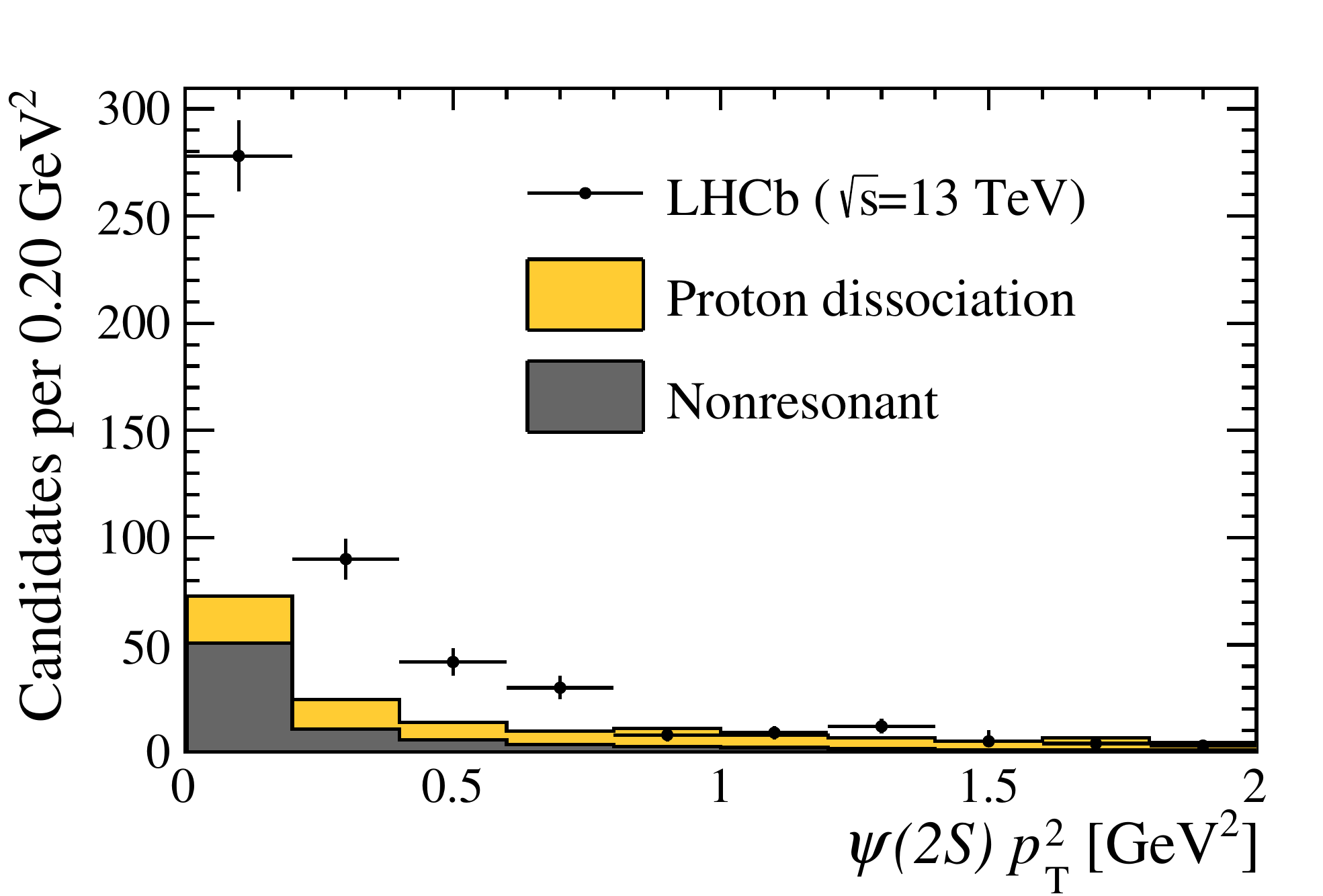}
\includegraphics[width=7.5cm]{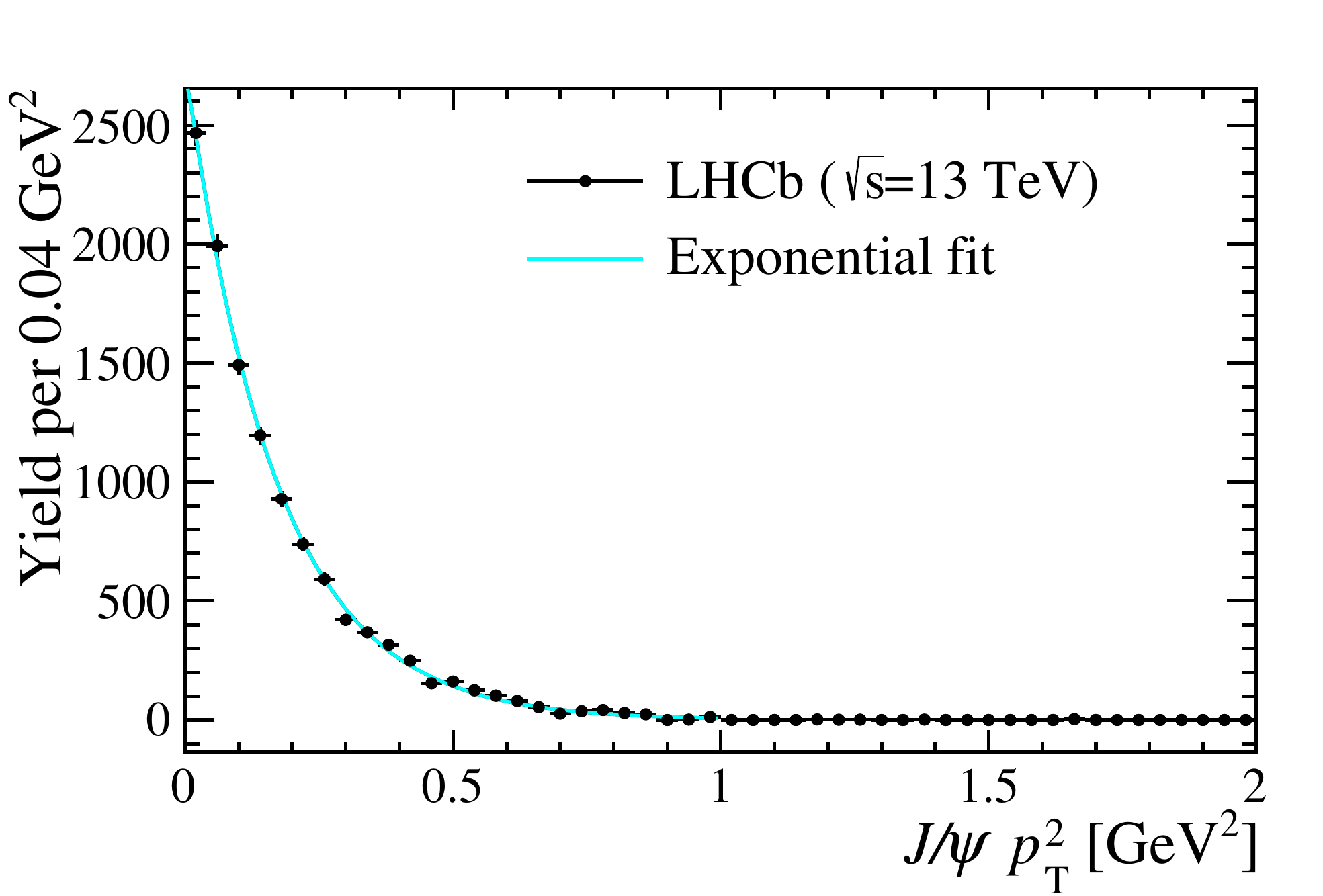}
\includegraphics[width=7.5cm]{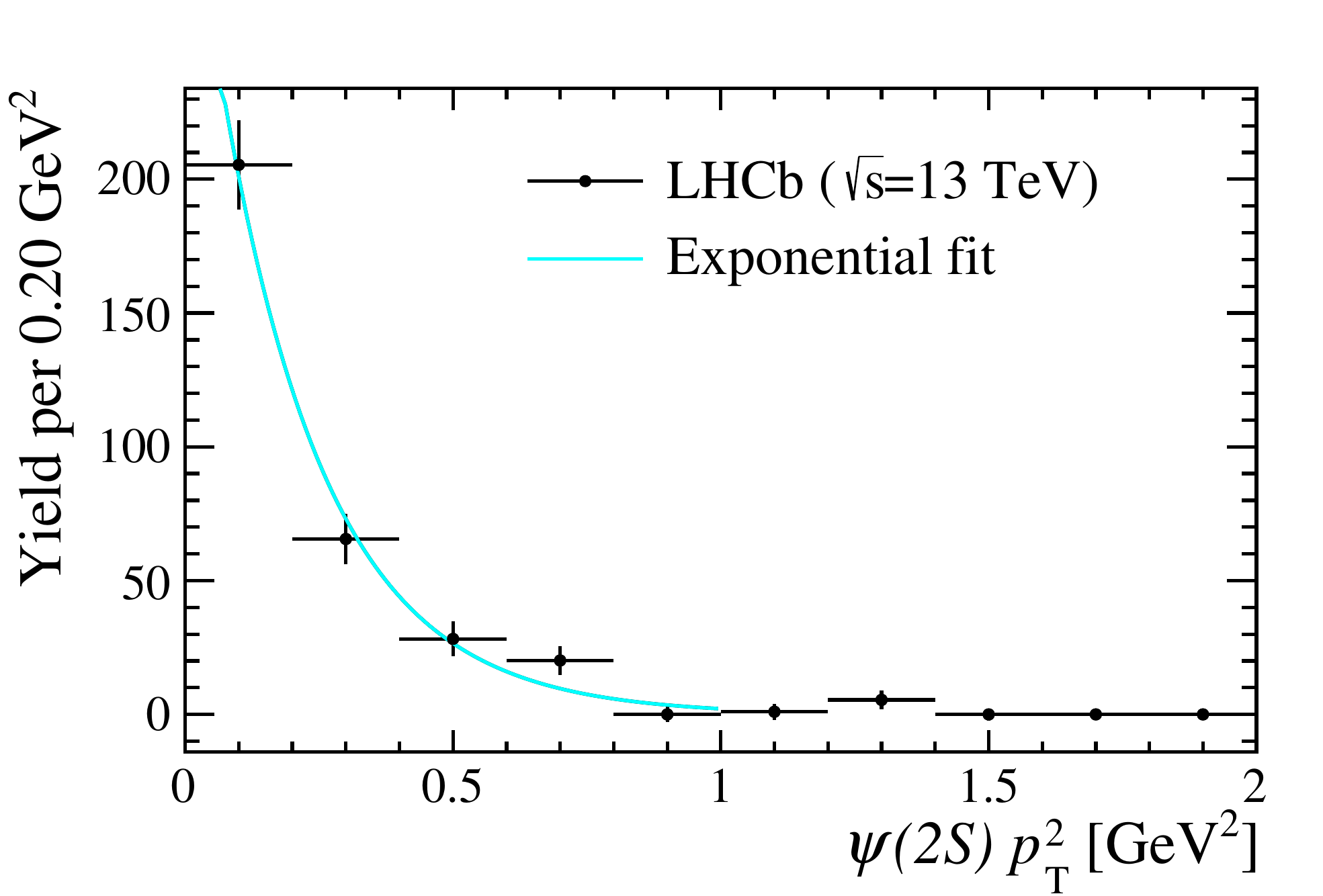}

\end{center}
\caption{Top: transverse momentum squared distribution of (left) \jpsi and (right) \psitwos candidates when data is below the \herschel threshold. Bottom: CEP signal for the (left) \jpsi and (right) \psitwos selections. The single exponential fit of the signal is shown by the curve 
superimposed on the data points. 
}
\label{fig:pt2_sel}
\end{figure}

The background subtracted distribution without \herschel veto applied is split into two distributions: $S_{\rm H}$ if
$\log(\chi^2_{\rm HRC})<3.5$ (corresponding to the signal selection), and $S_{\rm \bar{H}}$ otherwise.
Since $\epsilon_{\rm H}$ and $(1-\epsilon_{\rm H})$ are the respective efficiencies
for a CEP event to enter the distributions $S_{\rm H}$ and $S_{\rm \bar{H}}$, 
the distribution, $\beta=S_{\rm \bar{H}}-((1-\epsilon_{\rm H})/\epsilon_{\rm H})S_{\rm H}$, 
by construction has no contribution coming from exclusive events.
The distribution for $\beta$ approximates to the shape of the proton dissociation in 
the candidate distribution $S_{\rm H}$, 
but is not exactly the same since the efficiency to veto nonexclusive events has a weak dependence on $p_{\rm T}^2$.
Consequently,
the proton dissociation in the distribution $S_{\rm H}$ is estimated
by scaling the distribution $\beta$ by $f(p_{\rm T}^2)\equiv S_{\rm H}(p_{\rm T}^2)/\beta(p_{\rm T}^2)$.

 The scale factor $f(p_{\rm T}^2)$ is known from data for values of $p_{\rm T}^2 \gtrsim 0.8\gev^2$, since there is little signal in this region 
as the signal distribution is expected to follow $\exp(-b_{\rm sig} p_{\rm T}^2)$ with $b_{\rm sig}\approx 6\gev^{-2}$.
An extrapolation of $f(p_{\rm T}^2)$ is performed to the region $p_{\rm T}^2<0.8\gev^2$
using functions which fit the data well in the region $p_{\rm T}^2>0.8\gev^2$.
The default is an exponential function for the \jpsi analysis and a constant for the \psitwos analysis. A linear dependence is used to estimate the systematic uncertainty.

The $p_{\rm T}^2$ candidate distributions in data with the estimated backgrounds superimposed are shown
in the upper row of Fig.~\ref{fig:pt2_sel}. The lower row shows the signal components
after subtracting the proton dissociation background.
These are fitted with a single exponential function, $\exp(-b_{\rm sig}p_{\rm T}^2)$,
to test the hypothesis that the signal has this dependence.
The \jpsi signal contribution is well described with
$b_{\rm sig}=5.93\pm 0.08\gev^{-2}$,
consistent with extrapolations from previous $pp$ measurements at 7\tev and from H1
results~\cite{lhcb7,h1_latest}. The corresponding slope, in the \psitwos analysis, is $b_{\rm sig}=5.06\pm 0.45\gev^{-2}$.
Fits to the derived proton dissociation components show that these are also consistent
with a single exponential.

In the region $0<p_{\rm T}^2<0.8\gev^2$, $0.175\pm 0.015$ of the \jpsi candidate sample is estimated to be due to proton-dissociation events, while for the \psitwos sample the contamination is estimated to be $0.11\pm 0.06$. The uncertainties are statistical, and the correlation between the \herschel efficiency and the proton-dissociation contamination is taken into account.
The current analysis shows an approximate halving of the proton-dissociation background compared to the
analysis at $\sqrt{s}=7\tev$, due to the additional \herschel veto.
The overall purities are $0.755\pm0.015$ and $0.726\pm0.061$ for the \jpsi and \psitwos selections, respectively.

\subsection{Selection efficiency}
\label{sec:eff}

The efficiency for selecting signal events is the product of the reconstruction efficiency, $\epsilon_{\rm rec}$, and selection efficiency, $\epsilon_{\rm sel}$.  
The reconstruction efficiency is the product of trigger, tracking, muon chamber acceptance and muon
identification efficiencies. The acceptance is determined from simulation. The other quantities are determined from simulation and scaled
using a data calibration sample.
 The trigger efficiency is calibrated through the fraction of events where both muons pass the trigger, in a sample 
collected with the requirement that at least one muon passes the trigger.
The muon identification efficiency is calibrated using a sample enriched in \jpsi mesons
that has been selected requiring a single identified muon.
The tracking efficiency is calibrated using low-multiplicity events 
where the dimuon hardware was triggered by two objects having an absolute azimuthal angular difference close to $\pi$.

The efficiency for the selection requirements
on the mass and transverse momentum
of the \jpsi candidate, and the veto on additional tracks, photon activity, or \herschel
activity is obtained from data.

The fits to the mass distributions in Fig.~\ref{fig:mass} 
determine the fraction of signal inside the mass window and give a signal
efficiency of $0.967\pm0.002$. No dependence on rapidity is found.

The efficiency for the requirement on the meson candidates that
$p_{\rm T}^2<0.8\gev^2$ is $0.993\pm0.001$ and is determined from the 
fitted slope to the signal components shown in Fig.~\ref{fig:pt2_sel} as described in 
the previous section.
A small dependence on rapidity $y$ is introduced through the Regge extrapolation of the exponential slope:
$b=b_0+4\alpha^\prime\log(W/W_0)$, where $W^2=M_{\psi}e^{y}\sqrt{s}$.

The signal efficiency of vetoing events with additional VELO tracks or photons is 
obtained using the same technique described in Sec.~\ref{sec:hrceff} to determine the
\herschel veto efficiency.
When vetoing events with additional VELO tracks, no dependence on rapidity is found in simulation, while a slight dependence is observed for the photon veto, which is due to material effects in the detector whose density varies with rapidity.
The shape of the rapidity dependence is taken from simulation and normalised to data.
The efficiency of vetoing events with VELO tracks is determined to be 
$0.969\pm0.004$ and of vetoing events with photons is on average $0.983\pm0.003$.

\section{Cross-section calculation}
\label{sec:cs}

The products of the cross-sections and the branching fractions of the decays to two muons,
$\sigma_{\psi\rightarrow\mu\mu}$, are measured differentially in ten equally spaced bins of
\jpsi rapidity and three unequal bins of \psitwos rapidity in the range $y\in(2.0,4.5)$.
The measurements are limited to the fiducial region where both muons have pseudorapidities between 2.0 and 4.5.

The differential cross-section in each bin is

\begin{align}
  \label{eq:dsig}
\frac{\rm{d}\sigma_{\psi\rightarrow \mu^+ \mu^-}}{\rm{d}y}
(2.0<\eta_{\mu}<4.5)=
\frac{\mathcal{P}N}{\epsilon_{\rm rec}\epsilon_{\rm sel}\Delta y\epsilon_{\rm single}\mathcal{L}_{\rm tot}},
\end{align}
and the total cross-section, summed over all bins, is also calculated. In
Eq.~\ref{eq:dsig}, $N$ is the number of selected events, $\epsilon_{\rm rec}$ and $\epsilon_{\rm sel}$ are the efficiencies described in Sec.~\ref{sec:eff}, $\mathcal{P}$ is the purity given in Sec.~\ref{sec:pur},
$\Delta y$ is the width of the rapidity bin, 
$\mathcal{L}_{\rm tot}$ is the integrated luminosity
and
$\epsilon_{\rm single}$ is the efficiency for selecting single interaction
events, which accounts for the fact that the selection requirements reject
signal events that are accompanied by a visible proton-proton interaction
in the same beam crossing.

The number of visible $pp$ interactions per beam crossing, $v$, is assumed to follow a Poisson distribution, $P(v)=\mu^v e^{-\mu}/v!$.
The mean $\mu$ is determined from the fraction of beam crossings with no visible activity
and is calculated over the data-taking period in roughly hour-long intervals.  
The probability that a signal event is not rejected due to the presence
of another visible interaction 
is given by $P(0)$ and therefore $\epsilon_{\mathrm{single}}= e^{-\mu}$ which is equal to $0.3329 \pm 0.0003$. This
value is about 40\% higher
than the corresponding one
in the 7\tev analysis. The lower number of $pp$ interactions per
beam crossing at $\sqrt{s}=13\tev$ benefits the collection of CEP events. 
The integrated luminosity is evaluated as $204\pm8$\invpb  
and is found from $\mu$
and a constant of proportionality
that is measured in a dedicated calibration dataset~\cite{LHCb-PAPER-2014-047}.

\section{Systematic uncertainties}
\label{sec:sys}

Various sources of systematic uncertainties have been considered and
are summarised in Table~\ref{tab:sys} for the total cross-section.  Excluding the uncertainty on the luminosity,
they amount to 2.5\% in the \jpsi and 2.7\% in the \psitwos cases.

\begin{table}[!t]
\begin{center}
\caption{Summary of relative systematic uncertainties on the total cross-section.}
\label{tab:sys}
\begin{tabular}{lcc} 
 Source & \jpsi analysis (\%) & $\psi(2S)$ analysis (\%) \\ 
 \hline 
  \herschel veto & 1.7 & 1.7 \\ 
  2 VELO track & 0.2 & 0.2 \\ 
  0 photon veto & 0.2 & 0.2 \\ 
  Mass window & 0.6 & 0.6 \\ 
  $p_{\rm T}^2$ veto & 0.3 & 0.3 \\ 
  Proton dissociation & 0.7 & 0.7 \\  
  Feed-down & 0.7 & - \\ 
  Nonresonant & 0.1 & 1.5 \\ 
  Tracking efficiency & 0.7 & 0.7 \\ 
  Muon ID efficiency & 0.4 & 0.4 \\ 
  Trigger efficiency & 0.2 & 0.2 \\ 
 \hline 
 Total excluding luminosity& 2.5 & 2.7 \\ 
 \hline 
 Luminosity & 3.9 & 3.9 \\ 
 \hline 
\end{tabular}

\end{center}
\end{table}

The largest source of systematic uncertainty comes from the determination
  of the \herschel efficiency.
The fit to the $p^2_{\rm T}$ distribution in Fig.~\ref{fig:qedpt2} depends on 
assumptions made on the shape of the signal and background components. 
A systematic uncertainty is assessed firstly by changing the functional form of the background description,
secondly by fitting only the tail of the distribution and extrapolating the result to the signal, and thirdly
by using only the candidates in the first bin of the $p^2_{\rm T}$ distribution where the signal dominates. The differences of each to the nominal fit are combined in quadrature which results in a systematic uncertainty of 1.7\% on the total cross-section.

Since the same methodology is used to determine the efficiency for vetoing events with additional VELO tracks or photons, the associated systematic uncertainty is estimated with the same procedure. Since the simulation shows a dependence on rapidity for the efficiency due to the photon requirement, an additional uncertainty is added in quadrature in each rapidity bin, corresponding to the limited sample size of the simulation. This leads to a total systematic uncertainty of 0.2\% on the total cross-section due to each veto requirement.

The systematic uncertainty on the efficiency of the mass-window requirement is obtained by repeating the fit shown in Fig.~\ref{fig:mass} with the mass peak and resolution fixed to the values of the simulation. The fit is also repeated by changing the background description to a single exponential function across the whole region. The biggest difference with the nominal fit between these two alternative fits is taken as the systematic uncertainty, which is 0.6\% on the total cross-sections.

The uncertainty on the efficiency of selecting candidates with $p^2_{\rm T} <0.8\gev^2$ is 0.3\%. It is obtained by varying the signal shape from that shown in Fig.~\ref{fig:pt2_sel} to the one obtained by using the approach of the previous analysis \cite{lhcb7} where the $p^2_{\rm T}$ distribution is fitted with two exponential functions, one describing the proton dissociation and the other the signal shape. The slope and normalisations of each are free.
The difference in efficiency between the two approaches is added in quadrature to the 
uncertainty coming from the propagation of the uncertainties on the parameters describing
the Regge dependence that determines the rapidity dependence.

The proton-dissociation contamination depends on the extrapolation from the background-dominated high $p^2_{\rm T}$ region to the signal-dominated low $p^2_{\rm T}$ region.
The corresponding systematic uncertainty is assigned by changing the form of the extrapolation function from the nominal exponential one to an alternative linear function, or fitting the $p^2_{\rm T}$ distribution with two exponential functions to get the background contamination. The systematic uncertainty is the biggest difference between the nominal results and those from the two alternative approaches, and corresponds to 0.7\% on the total cross-section.

The systematic uncertainty due to the feed-down contribution in the \jpsi analysis 
is assessed to be 0.7\% on the total cross-section. It corresponds to 
the largest difference in the cross-section determination
from a series of alternative fits to the $\jpsi+\gamma$ spectrum
in which the photon energy scale, photon detection efficiency,
invariant mass resolution, material interactions, and the \psitwos contribution,
are each varied by their estimated uncertainties.

An alternative estimate of the nonresonant background in 
Fig.~\ref{fig:mass} is performed by fitting
a single exponential function between 1.5 and 2.5\gev and 
extrapolating this into the signal region.
This changes the total cross-section by
0.1\% in the \jpsi analysis and 1.5\% in the \psitwos analysis.
These values are taken as systematic uncertainties due to the nonresonant background.

The reconstruction efficiency is taken from simulated events and calibrated using data.
The technique depends on tagging a muon that fired the trigger
and probing a partially reconstructed track that forms a \jpsi candidate.
To assess the systematic uncertainty due to the method, this technique is applied to two simulated samples that have different tracking efficiencies.
The resulting tracking efficiencies are compared after calibration using data.
In a second test of the methodology, one simulated sample is taken as pseudodata and the other simulated sample applies the calibration procedure.
The resulting efficiencies are compared to the true values in the pseudodata.
The largest
difference in each rapidity bin is assigned as a systematic  uncertainty, which is assumed 
to be fully correlated between bins, and varies from 0.5\% to 3.1\% depending on the sample size. 
A systematic uncertainty on the method used in evaluating the muon identification
and trigger efficiencies is assigned by comparing the derived values 
in simulation with truth, resulting in a 0.4\% uncertainty on the total cross-section due
to the muon identification,
and 0.2\% due to the trigger. The systematic uncertainty on the muon chamber acceptance is determined from the difference in the kinematic distributions in data and simulation, and its effect on the final reconstruction efficiency systematic uncertainty is negligible in all bins.

A bin migration uncertainty has been estimated
using simulation to relate the reconstructed and true rapidity bin. 
The difference is smaller than 0.06\% in all bins and so is considered negligible.

 Most systematic uncertainties are assumed to be 100\% correlated between rapidity bins except the photon-veto-shape systematic uncertainty, which is assumed to be independent between bins as it depends on the statistical precision of the simulation. 
As the determination of the sample purity depends on the \herschel efficiency, these two quantitities are correlated. The correlation factors are determined in simulation, and the values are $\rho=-0.50$ and $\rho=-0.06$ for the ${J/\psi}$ and \psitwos selection, respectively. The lower statistical
precision of the \psitwos sample imposes less constraint on the proton dissociation scale factor $f(p_{\rm T}^2)$ and results in a smaller correlation.  
The total systematic uncertainties are given in Table~\ref{tab:sys} taking account of the correlations.


\section{Results}
\label{sec:results}


The product of the differential cross-sections
and branching fractions to two muons, with both
 muons inside the fiducial acceptance $2.0<\eta<4.5$, 
are given per meson rapidity bin in Tables~\ref{tab:num} and \ref{tab:numpsi_sig} for \jpsi and \psitwos mesons, respectively. The tables also present a summary of the 
 numbers entering the cross-section calculation.
\pagebreak

\begin{table}[!t]
\vspace{1cm}
  \centering
\caption{Tabulation of numbers entering the cross-section calculation for the \jpsi analysis with statistical and systematic uncertainties for the integrated luminosity of $\mathcal{L}_{\rm tot}=204\pm8$\invpb and the fraction of single-interaction beam crossings, $\epsilon_{\rm single}=0.3329 \pm 0.0003$.}
\label{tab:num}
\begin{tabular}{lccccc} 
 $y$ bin &{ 2.0$-$2.25} & {2.25$-$2.5} & {2.5$-$2.75}& {2.75$-$3.0} & {3.0$-$3.25} \\ 
\hline  
 \hline 
 $N$ & 259& 1022& 1644& 2204& 2482\\ 
 Stat. unc. (\%)& 6.2& 3.1& 2.5& 2.1& 2.0\\ 
 \hline 
  $\epsilon_{\rm rec}$ & 0.410 & 0.525 & 0.555 & 0.565 & 0.563 \\ 
 Stat. unc. (\%) & 5.9 & 4.2 & 3.3 & 2.8 & 2.6 \\ 
 Syst. unc. (\%) & 3.1 & 0.8 & 1.7 & 1.0 & 0.5 \\ 
 \hline 
 $\epsilon_{\rm sel}$ & 0.636 & 0.643 & 0.650 & 0.655 & 0.663 \\ 
 Stat. unc. (\%) & 1.2 & 1.2 & 1.2 & 1.2 & 1.2 \\ 
 Syst. unc. (\%) & 2.5 & 2.0 & 2.0 & 1.9 & 1.9 \\ 
 \hline 
 Purity  &  0.760 &  0.759 &  0.751 &  0.758 &  0.764 \\ 
 Stat. unc. (\%)  &  2.7 &  2.2 &  2.2 &  2.1 &  2.1 \\ 
 Syst. unc. (\%)  &  1.0 &  1.0 &  1.0 &  1.0 &  1.0 \\ 
 \hline 
 $\mathrm{d}\sigma/\mathrm{d}y$(pb) & 44 & 134 & 200 & 263 & 296 \\ 
 Stat. unc. (\%) & 9.2 & 6.0 & 5.0 & 4.5 & 4.3 \\ 
 Syst. unc. (\%) & 4.3 & 2.7 & 3.1 & 2.7 & 2.6 \\ 
 Lumi. unc. (\%) & 3.9 & 3.9 & 3.9 & 3.9 & 3.9 \\
\hline 
 & \multicolumn{5}{c}{ }\\
 
  $y$ bin &{ 3.25$-$3.50} & {3.50$-$3.75} & {3.75$-$4.0}& {4.0$-$4.25} & {4.25$-$4.5} \\ 
\hline  
 \hline 
 $N$ & 2522& 2112& 1433& 829& 246\\ 
 Stat. unc. (\%)& 2.0& 2.2& 2.6& 3.5& 6.4\\ 
 \hline 
  $\epsilon_{\rm rec}$ & 0.587 & 0.599 & 0.588 & 0.551 & 0.518 \\ 
 Stat. unc. (\%) & 2.5 & 2.6 & 2.8 & 3.3 & 4.1 \\ 
 Syst. unc. (\%) & 0.6 & 0.6 & 0.5 & 0.8 & 0.9 \\ 
 \hline 
 $\epsilon_{\rm sel}$ & 0.665 & 0.670 & 0.670 & 0.676 & 0.667 \\ 
 Stat. unc. (\%) & 1.2 & 1.2 & 1.2 & 1.2 & 1.2 \\ 
 Syst. unc. (\%) & 1.9 & 1.9 & 1.9 & 1.9 & 2.0 \\ 
 \hline 
 Purity  &  0.763 &  0.749 &  0.748 &  0.732 &  0.738 \\ 
 Stat. unc. (\%)  &  2.1 &  2.1 &  2.2 &  2.4 &  3.1 \\ 
 Syst. unc. (\%)  &  1.0 &  1.0 &  1.0 &  1.0 &  1.0 \\ 
 \hline 
  $\mathrm{d}\sigma/\mathrm{d}y$(pb) & 288 & 230 & 159 & 95 & 31 \\ 
 Stat. unc. (\%) & 4.3 & 4.4 & 4.8 & 5.7 & 8.5 \\ 
 Syst. unc. (\%) & 2.6 & 2.6 & 2.6 & 2.7 & 2.8 \\ 
 Lumi. unc. (\%) & 3.9 & 3.9 & 3.9 & 3.9 & 3.9 \\ 
 \hline 
 \end{tabular}

\vspace{3cm}
\end{table}

\pagebreak

\begin{table}[!t]
\begin{center}
  \caption{Tabulation of numbers entering the cross-section calculation for the \psitwos analysis with statistical and systematic uncertainties for the integrated luminosity of $\mathcal{L}_{\rm tot}=204\pm8$\invpb and the fraction of single-interaction beam crossings, $\epsilon_{\rm single}=0.3329 \pm 0.0003$.}

\label{tab:numpsi_sig}
\begin{tabular}{lccc} 
 $y$ bin &{ 2.0$-$3.0} & {3.0$-$3.5} & {3.5$-$4.5}  \\ 
\hline  
 \hline 
 $N$ & 170& 134& 136\\ 
 Stat. unc. (\%)& 7.7& 8.6& 8.6\\ 
 \hline 
  $\epsilon_{\rm rec}$ & 0.633 & 0.644 & 0.622 \\ 
 Stat. unc. (\%) & 3.4 & 2.6 & 2.9 \\ 
 Syst. unc. (\%) & 1.3 & 0.6 & 0.6 \\ 
 \hline 
 $\epsilon_{\rm sel}$ & 0.650 & 0.664 & 0.671 \\ 
 Stat. unc. (\%) & 1.2 & 1.2 & 1.2 \\ 
 Syst. unc. (\%) & 1.9 & 1.9 & 1.9 \\ 
 \hline 
 Purity & \multicolumn{3}{c}{ 0.726 }\\ 
 Stat. unc. (\%)  & \multicolumn{3}{c}{ 8.4 }\\ 
 Syst. unc. (\%)  & \multicolumn{3}{c}{ 1.7 }\\ 
 \hline 
 $\mathrm{d}\sigma/\mathrm{d}y$(pb) & 4.4 & 6.6 & 3.4 \\ 
 Stat. unc. (\%) & 12.0 & 12.4 & 12.4 \\ 
 Syst. unc. (\%) & 2.9 & 2.7 & 2.7 \\ 
 Lumi. unc. (\%) & 3.9 & 3.9 & 3.9 \\ 
 \hline 
 \end{tabular}

\end{center}

  \small
  \caption{
Tabulation, in bins of meson rapidity, of the fraction of decays with
both muons in the range $2.0<\eta<4.5$ and the differential cross-sections for
\jpsi and \psitwos production calculated without fiducial requirements on the muons.}
  \label{tab:acccs}
  \begin{center}
    \begin{tabular}{lccccc} 
 \jpsi $y$ bin &{ 2.0$-$2.25} & {2.25$-$2.5} & {2.5$-$2.75}& {2.75$-$3.0} & {3.0$-$3.25} \\ 
 \hline 
 Acc. & $0.095 \pm 0.003$& $0.280 \pm 0.005$& $0.460 \pm 0.006$& $0.627 \pm 0.006$& $0.733 \pm 0.005$\\ 
 $\frac{\mathrm{d}\sigma}{\mathrm{d}y}$(nb) & $7.76 \pm 0.77$& $8.03 \pm 0.51$& $7.29 \pm 0.38$& $7.04 \pm 0.33$& $6.78 \pm 0.30$\\ 
 \hline  
 & & & & &\\
 
 
 \jpsi $y$ bin &{ 3.25$-$3.50} & {3.50$-$3.75} & {3.75$-$4.0}& {4.0$-$4.25} & {4.25$-$4.5} \\ 
 \hline 
 Acc. & $0.721 \pm 0.005$& $0.620 \pm 0.006$& $0.471 \pm 0.006$& $0.287 \pm 0.006$& $0.094 \pm 0.004$\\ 
$\frac{\mathrm{d}\sigma}{\mathrm{d}y}$(nb) & $6.70 \pm 0.29$& $6.22 \pm 0.28$& $5.66 \pm 0.29$& $5.55 \pm 0.34$& $5.46 \pm 0.52$\\ 
 \hline 
 \end{tabular}

  \end{center}
  \begin{center}
    \begin{tabular}{lccc} 
$\psi(2S)$ $y$ bin &{ 2.0$-$3.0} & {3.0$-$3.5} & {3.5$-$4.5}  \\ 
 \hline 
 Acc. & $0.362 \pm 0.003$& $0.726 \pm 0.004$& $0.372 \pm 0.003$\\ 
  $\frac{\mathrm{d}\sigma}{\mathrm{d}y}$(nb) & $1.53 \pm 0.25$& $1.16 \pm 0.19$& $1.17 \pm 0.20$\\ 
 \hline 
 \end{tabular}

  \end{center}
\end{table}

 The correlations between the statistical and systematic uncertainties
in each bin are shown in Tables \ref{tab:corr_sig} and \ref{tab:corr_sigpsi} in the Appendix.
Summing these differential results leads to measurements of the product of the cross-sections and branching fractions, 
where both muons are within the fiducial region, $2.0<\eta<4.5$:

$$
 \begin{array}{rcl}
\sigma_{J/\psi\rightarrow\mu^+\mu^-}(2<\eta<4.5)&=&435 \pm 18 \pm 11 \pm 17 {\rm \ pb}\\ 
\sigma_{\psi(2S)\rightarrow\mu^+\mu^-}(2<\eta<4.5)&=&11.1 \pm 1.1 \pm 0.3 \pm 0.4 {\rm \ pb}.\\ 
\end{array}
 $$ 
The first uncertainties are statistical and include the uncertainties on the data-driven efficiencies and purities,
the second are systematic, 
and the third are due to the luminosity determination.

As a cross-check and to confirm the improvements brought by \herschel, the cross-sections have been recalculated without imposing the \herschel veto: consistent results are obtained but with a larger systematic uncertainty of about 8\%.
While the extracted signal contribution is comparable to Fig.~\ref{fig:pt2_sel} and well described by a
single exponential function with a consistent value of $b_{\rm sig}=5.92 \pm 0.06\gev^{-2}$, the extracted
proton-dissociation component requires two exponential functions to describe the distribution.

\begin{figure}[!t]
\begin{center}
\includegraphics[width=10cm]{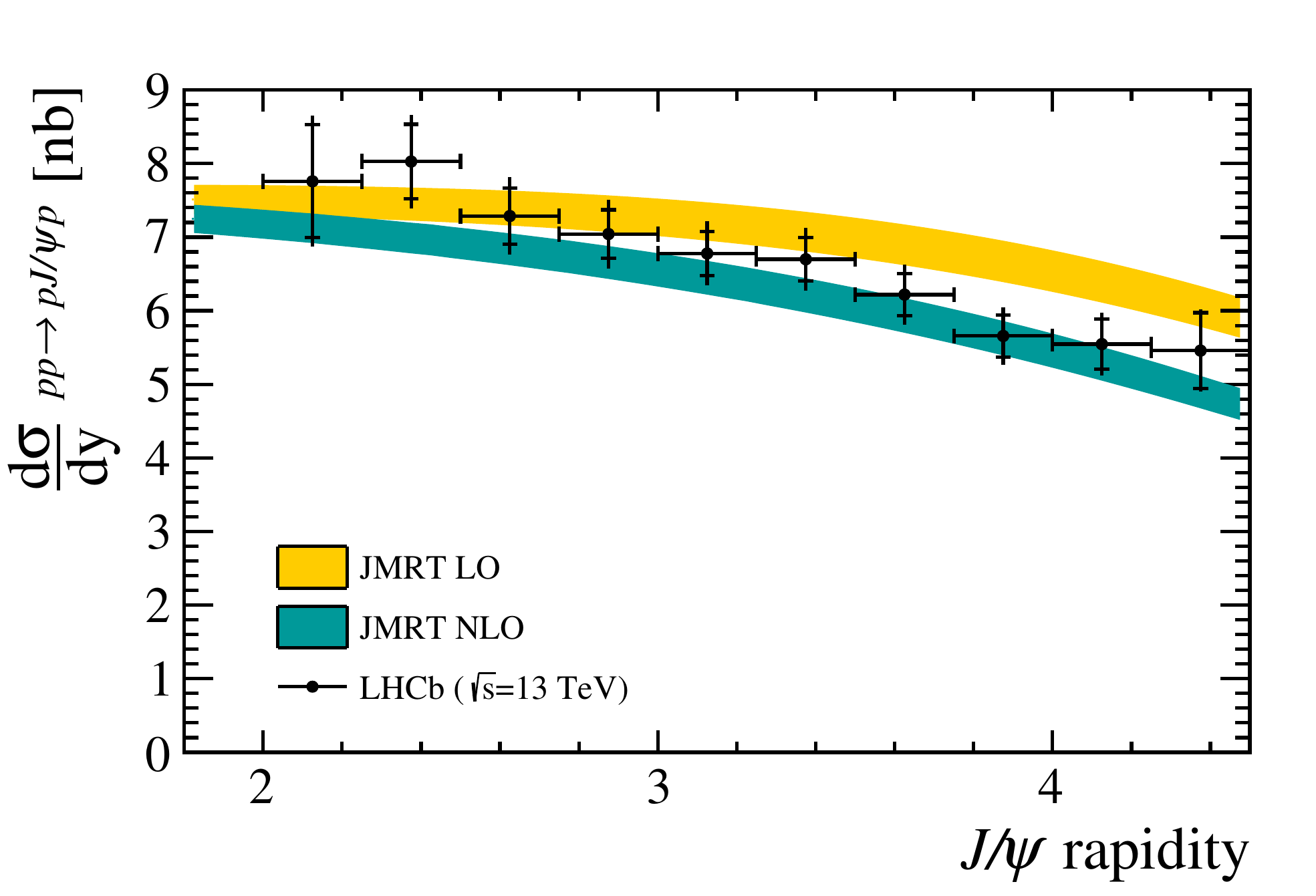}
\includegraphics[width=10cm]{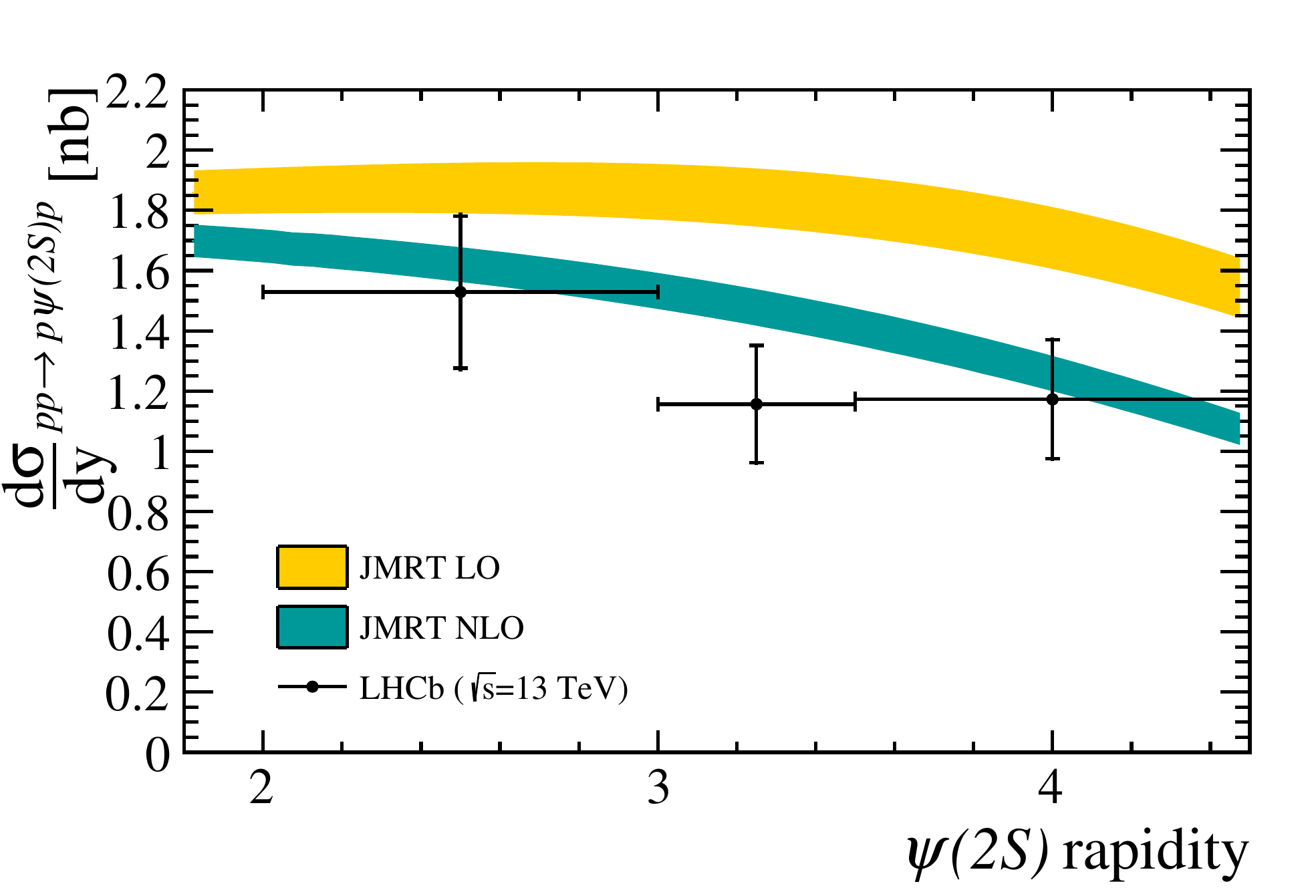}
\end{center}
\caption{Differential cross-sections compared to LO and NLO theory JMRT predictions~\cite{jones,jonespsi}
for the \jpsi meson (top) and the \psitwos meson (bottom).
The inner error bar represents the statistical uncertainty; the outer is the total uncertainty. Since the systematic uncertainty for the \psitwos meson is negligible with respect to the statistical uncertainty, it is almost not visible in the lower figure.} 
\label{fig:difcs}
\end{figure}

To compare with theoretical predictions, which are generally expressed without
fiducial requirements on the muons, 
the differential cross-sections for \jpsi and \psitwos mesons as functions of the meson rapidity are calculated
by correcting for the branching
fractions to muon pairs, 
$\mathcal{B}(J/\psi\rightarrow\mu^+\mu^-)=(5.961\pm0.033)$\% and $\mathcal{B}(\psi(2S)\rightarrow\mu^+\mu^-)=(0.79\pm0.09)$\% ~\cite{PDG2017}, and for the fraction of those muons that fall inside the fiducial acceptance of the measurement.
The fiducial acceptance is determined using SuperCHIC~\cite{superchic} assuming that the polarisation of the meson is the same as that of the photon.
The acceptance values in bins of meson rapidity
are tabulated in Table~\ref{tab:acccs} along with the 
differential cross-section results.  
These are plotted in Fig.~\ref{fig:difcs} and compared to
the theoretical calculations of Refs.~\cite{jones,jonespsi}.  
Both measurements are in better agreement with the next-to-LO (NLO) predictions.
The $\chi^2/$ndf for the \jpsi analysis
is 8.1/10 while for the \psitwos analysis, it is 3.0/3. They are less consistent with the LO predictions having
28.5/10 and 11.0/3 for the \jpsi and \psitwos analysis, respectively.

The cross-section for the CEP of vector mesons in $pp$ collisions
is related to the photoproduction cross-section, 
$\sigma_{\gamma p\rightarrow \psi p}$~\cite{jones},
\begin{equation}
\sigma_{pp\rightarrow p\psi p} = 
r(W_+ )k_+{\mathrm{d}n\over \mathrm{d}k_+} \sigma_{\gamma p\rightarrow \psi p} (W_+)   +
r(W_-) k_-{\mathrm{d}n\over \mathrm{d}k_-} \sigma_{\gamma p\rightarrow \psi p} (W_-).
\label{eq:photo}
\end{equation}
Here, $r$ is the gap survival factor,
$k_\pm\equiv M_\psi/2 e^{\pm y}$ is the photon energy, $\mathrm{d}n/\mathrm{d}k_\pm$ is the photon flux
and $W_\pm^2=2k_\pm\sqrt{s}$ is     
the invariant mass of the photon-proton system.
Equation \ref{eq:photo} shows that there is a two-fold ambiguity with $W_+,W_-$ both contributing to one LHCb rapidity bin.
Since the $W_-$ solution contributes about one third and as it has been 
previously measured at HERA, this term is fixed 
using the H1 parametrisation of their results~\cite{h1_latest}:
$\sigma_{\gamma p\rightarrow J/\psi p}=a(W/90{\rm\gev})^\delta$ with $a=81\pm 3$ pb and
$\delta=0.67\pm0.03$.
For the \psitwos $W_-$ solution, the H1 \jpsi parametrisation is scaled by 0.166, their measured ratio of
\psitwos to \jpsi cross-sections~\cite{h1psi}.
The photon flux is taken from Ref.\cite{kepka} and
the gap survival probabilities are taken from Ref.~\cite{Jones:2016icr}.
With these inputs, 
which for ease of calculation
are reproduced in Tables~\ref{tab:gap} and \ref{tab:gap7} in the Appendix,
Eq.~\ref{eq:photo} allows the calculation of 
$\sigma_{\gamma p\rightarrow \psi p}$ at high values of $W$ beyond the kinematic reach of HERA.

\begin{figure}[!t]
  \begin{center}
    \includegraphics[width=10cm]{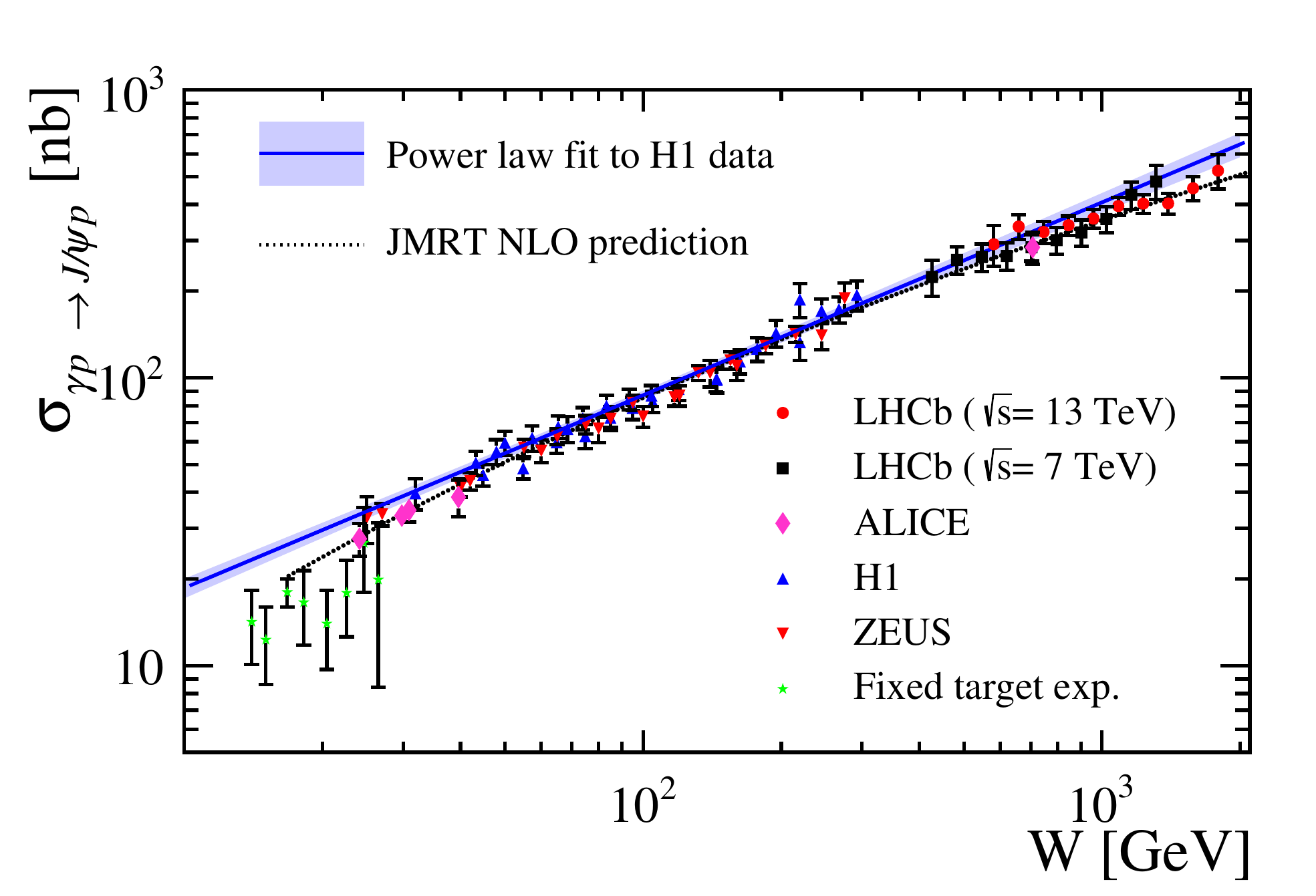}
    \includegraphics[width=10cm]{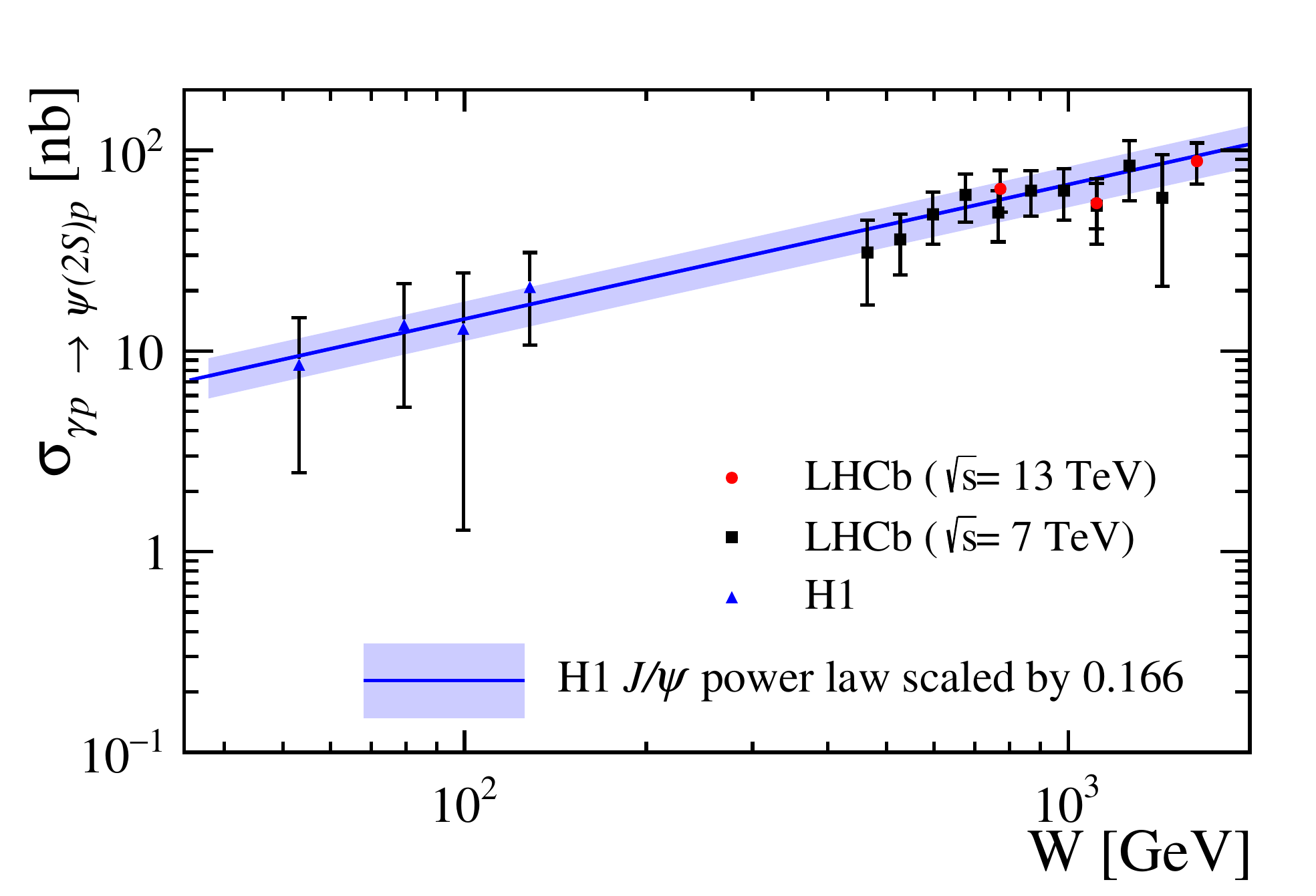}    
  \end{center}
  \caption{Compilation of photoproduction cross-sections for various experiments. 
The upper (lower) plot uses the \jpsi (\psitwos) data. 
}
\label{fig:photop}
\end{figure}

The photoproduction cross-sections for \jpsi and \psitwos
are shown in Fig.~\ref{fig:photop}.
It includes a comparison to H1~\cite{h1_latest}, ZEUS~\cite{zeus_jpsi} and ALICE~\cite{alice} results, 
and at lower $W$ values fixed target data from ~E401~\cite{e401}, E516~\cite{e516} and E687~\cite{e687}.
Also shown are previous LHCb results
at $\sqrt{s}=7\tev$, recalculated using improved photon flux and gap survival factors. The 13\tev LHCb data are in agreement with the 7\tev results in the kinematic region where they
overlap. However, the 13\tev data extends the $W$ reach to almost 2\tev.
Figure~\ref{fig:photop} also shows the power-law fit to H1 data~\cite{h1_latest} and it 
can be seen that this is insufficient 
to describe the {\jpsi} data at the highest energies.
In contrast, the data is in good agreement with the JMRT prediction,
which takes account of most of the NLO QCD effects~\cite{Jones:2016icr} and deviates from a simple
power-law shape at high $W$.

\section{Conclusions}
\label{sec:conclude}

Measurements
are presented of the cross-sections times branching fractions
for exclusive \jpsi and \psitwos mesons decaying to muons
with pseudorapidities between 2.0 and 4.5.
The addition of new scintillators in the forward region
has resulted in lower backgrounds
in $pp$ collisions at a centre-of-mass energy $\sqrt{s}=13\tev$ compared to the previous measurement at $\sqrt{s}=7\tev$. 
As a consequence, the systematic uncertainty on the \jpsi cross-section is reduced from 5.6\% at $\sqrt{s}=7\tev$ to 2.7\% at $\sqrt{s}=13\tev$, reflecting an improved understanding of the background proton-dissociation process.
After correcting for the muon acceptance, the cross-sections for the
\jpsi and \psitwos mesons are compared to theory and found to be in better agreement
with the JMRT NLO rather than LO predictions.  
The derived cross-section for \jpsi photoproduction shows
a deviation from a pure power-law extrapolation of H1 data,
while the \psitwos results are consistent although
more data are required in this channel to make a critical comparison.

\section*{Acknowledgements}
%
%
\noindent We express our gratitude to our colleagues in the CERN
accelerator departments for the excellent performance of the LHC. We
thank the technical and administrative staff at the LHCb
institutes. We acknowledge support from CERN and from the national
agencies: CAPES, CNPq, FAPERJ and FINEP (Brazil); MOST and NSFC
(China); CNRS/IN2P3 (France); BMBF, DFG and MPG (Germany); INFN
(Italy); NWO (Netherlands); MNiSW and NCN (Poland); MEN/IFA
(Romania); MinES and FASO (Russia); MinECo (Spain); SNSF and SER
(Switzerland); NASU (Ukraine); STFC (United Kingdom); NSF (USA).  We
acknowledge the computing resources that are provided by CERN, IN2P3
(France), KIT and DESY (Germany), INFN (Italy), SURF (Netherlands),
PIC (Spain), GridPP (United Kingdom), RRCKI and Yandex
LLC (Russia), CSCS (Switzerland), IFIN-HH (Romania), CBPF (Brazil),
PL-GRID (Poland) and OSC (USA). We are indebted to the communities
behind the multiple open-source software packages on which we depend.
Individual groups or members have received support from AvH Foundation
(Germany), EPLANET, Marie Sk\l{}odowska-Curie Actions and ERC
(European Union), ANR, Labex P2IO and OCEVU, and R\'{e}gion
Auvergne-Rh\^{o}ne-Alpes (France), Key Research Program of Frontier
Sciences of CAS, CAS PIFI, and the Thousand Talents Program (China),
RFBR, RSF and Yandex LLC (Russia), GVA, XuntaGal and GENCAT (Spain),
Herchel Smith Fund, the Royal Society, the English-Speaking Union and
the Leverhulme Trust (United Kingdom).


\addcontentsline{toc}{section}{References}
\setboolean{inbibliography}{true}
\bibliographystyle{LHCb}
\bibliography{main,main1,LHCb-PAPER,LHCb-DP,LHCb-TDR}

\pagebreak
\centerline{\large\bf LHCb collaboration}
\begin{flushleft}
\small
R.~Aaij$^{27}$,
B.~Adeva$^{41}$,
M.~Adinolfi$^{48}$,
C.A.~Aidala$^{73}$,
Z.~Ajaltouni$^{5}$,
S.~Akar$^{59}$,
P.~Albicocco$^{18}$,
J.~Albrecht$^{10}$,
F.~Alessio$^{42}$,
M.~Alexander$^{53}$
% ...

{\footnotesize \it
% ...
$ ^{5}$Clermont Universit{\'e}, Universit{\'e} Blaise Pascal, CNRS/IN2P3, LPC, Clermont-Ferrand, France\\
% ...
$ ^{10}$Fakult{\"a}t Physik, Technische Universit{\"a}t Dortmund, Dortmund, Germany\\
% ...
$ ^{18}$INFN Laboratori Nazionali di Frascati, Frascati, Italy\\
% ...
$ ^{27}$Nikhef National Institute for Subatomic Physics, Amsterdam, Netherlands\\
% ...
$ ^{41}$Instituto Galego de F{\'\i}sica de Altas Enerx{\'\i}as (IGFAE), Universidade de Santiago de Compostela, Santiago de Compostela, Spain\\
$ ^{42}$European Organization for Nuclear Research (CERN), Geneva, Switzerland\\
% ...
$ ^{48}$H.H. Wills Physics Laboratory, University of Bristol, Bristol, United Kingdom\\
% ...
$ ^{53}$School of Physics and Astronomy, University of Glasgow, Glasgow, United Kingdom\\
% ...
$ ^{59}$University of Cincinnati, Cincinnati, OH, United States\\
% ...
$ ^{61}$Syracuse University, Syracuse, NY, United States\\
% ...
$ ^{73}$University of Michigan, Ann Arbor, United States, associated to $^{61}$\\
% ...
}
\end{flushleft}
 
\newpage

\appendix
\section{Additional material}

\begin{table}[!h]
\small
\begin{center}
\caption{(Top) Statistical and (bottom) systematic correlation matrices for \jpsi, where each column corresponds to one rapidity bin in increasing order. As the matrix is symmetric, only the top triangle is shown.}
\label{tab:corr_sig}
\begin{tabular}{cccccccccc} 
  1.00& 0.58& 0.58& 0.57& 0.57& 0.57& 0.56& 0.54& 0.51& 0.40 \\ 
 & 1.00& 0.71& 0.71& 0.71& 0.71& 0.70& 0.67& 0.62& 0.49 \\ 
 & & 1.00& 0.74& 0.74& 0.74& 0.73& 0.69& 0.64& 0.50 \\ 
 & & & 1.00& 0.76& 0.75& 0.74& 0.71& 0.65& 0.50 \\ 
 & & & & 1.00& 0.76& 0.74& 0.71& 0.65& 0.50 \\ 
 & & & & & 1.00& 0.74& 0.71& 0.65& 0.50 \\ 
 & & & & & & 1.00& 0.69& 0.64& 0.49 \\ 
 & & & & & & & 1.00& 0.61& 0.46 \\ 
 & & & & & & & & 1.00& 0.43 \\ 
 & & & & & & & & & 1.00 \\ 
\end{tabular}

\begin{center}
\begin{tabular}{cccccccccc} 
  1.00& 0.74& 0.88& 0.78& 0.68& 0.69& 0.71& 0.64& 0.77& 0.76 \\ 
& 1.00& 0.91& 0.96& 0.95& 0.95& 0.96& 0.94& 0.96& 0.93 \\ 
& & 1.00& 0.94& 0.88& 0.89& 0.91& 0.86& 0.94& 0.92 \\ 
& & & 1.00& 0.97& 0.97& 0.98& 0.95& 0.98& 0.95 \\ 
 & & & & 1.00& 0.99& 0.99& 0.99& 0.97& 0.93 \\ 
 & & & & & 1.00& 0.99& 0.98& 0.98& 0.94 \\ 
 & & & & & & 1.00& 0.99& 0.99& 0.95 \\ 
 & & & & & & & 1.00& 0.97& 0.93 \\ 
 & & & & & & & & 1.00& 0.96 \\ 
 & & & & & & & & & 1.00 \\ 
\end{tabular}

\end{center}
\begin{center}
\caption{(Top) Statistical and (bottom) systematic correlation matrices for \psitwos, where each column corresponds to one rapidity bin in increasing order. As the matrix is symmetric, only the top triangle is shown.}
\label{tab:corr_sigpsi}
\begin{tabular}{ccc} 
  1.00& 0.55& 0.56 \\ 
& 1.00& 0.52 \\ 
& & 1.00 \\ 
\end{tabular}

\end{center}
\centering
\begin{tabular}{ccc} 
  1.00& 0.95& 0.96 \\ 
& 1.00& 1.00 \\ 
& & 1.00 \\ 
\end{tabular}

\end{center}
\end{table}

\begin{table}[!h]
  \centering
  \caption{Values used in evaluating the photo-production cross-section using Eq.~\ref{eq:photo}
  for the \jpsi and \psitwos analysis with gap survival factors for the production of \jpsi and \psitwos mesons at $\sqrt{s}=13$~TeV~\cite{Jones:2016icr}. For the $J/\psi$ analyis, $\sigma_{\gamma p\rightarrow J/\psi p(W_+)}$ is calculated using the power-law description of HERA or the JMRT NLO description for $\sigma_{\gamma p\rightarrow J/\psi p(W_-)}$.  }
  \label{tab:gap}
\begin{tabular}{lccccc} 
 \jpsi $y$ bin &{ 2.0$-$2.25} & {2.25$-$2.5} & {2.5$-$2.75}& {2.75$-$3.0} & {3.0$-$3.25} \\ 
 \hline 
 $W_+$ (GeV)& 581& 658& 746& 845& 958 \\ 
 $k_+ \mathrm{d}n/\mathrm{d}k_+(\times 10^{-3})$& 22.7& 21.6& 20.4& 19.2& 18.0 \\ 
 $r(W_+)$& 0.786& 0.774& 0.762& 0.748& 0.732 \\ 
 $W_-$ (GeV)& 69.4& 61.2& 54.0& 47.7& 42.1 \\ 
 $k_- \mathrm{d}n/\mathrm{d}k_-(\times 10^{-3})$& 42.5& 43.7& 44.9& 46.0& 47.2 \\ 
 $r(W_-)$& 0.885& 0.888& 0.891& 0.893& 0.896 \\ 
 $\sigma_{\gamma p\rightarrow J/\psi p(W_-)}$ (nb) & & & & &  \\ 
 Power law & 68.0& 62.6& 57.6& 52.9& 48.7 \\ 
 JMRT NLO & 65.3& 59.5& 54.1& 49.1& 44.5 \\ 
 \hline 
 Calculated: & & & & &  \\ 
 $\sigma_{\gamma p\rightarrow J/\psi p(W_+)}$ (nb) & & & & &  \\ 
 Power Law & 291& 335& 321& 339& 358 \\ 
 JMRT NLO & 297& 343& 330& 350& 371 \\ 
 \hline 
 & & & & &  \\ 

 \jpsi $y$ bin &{ 3.25$-$3.50} & {3.50$-$3.75} & {3.75$-$4.0}& {4.0$-$4.25} & {4.25$-$4.5} \\ 
 \hline 
 $W_+$ (GeV)& 1085& 1230& 1394& 1579& 1790 \\ 
 $k_+ \mathrm{d}n/\mathrm{d}k_+(\times 10^{-3})$& 16.8& 15.7& 14.5& 13.3& 12.1 \\ 
 $r(W_+)$& 0.715& 0.695& 0.672& 0.647& 0.618 \\ 
 $W_-$ (GeV)& 37.1& 32.8& 28.9& 25.5& 22.5 \\ 
 $k_- \mathrm{d}n/\mathrm{d}k_-(\times 10^{-3})$& 48.3& 49.5& 50.7& 51.8& 53.0 \\ 
 $r(W_-)$& 0.898& 0.901& 0.903& 0.905& 0.907 \\ 
 $\sigma_{\gamma p\rightarrow J/\psi p(W_-)}$ (nb) & & & & &  \\ 
 Power law & 44.8& 41.2& 37.9& 34.8& 32.0 \\ 
 JMRT NLO & 40.2& 36.3& 32.7& 29.5& 26.4 \\ 
 \hline 
 Calculated: & & & & &  \\ 
 $\sigma_{\gamma p\rightarrow J/\psi p(W_+)}$ (nb) & & & & &  \\ 
Power Law & 395& 403& 403& 456& 524 \\ 
 JMRT NLO & 411& 423& 427& 485& 560 \\ 
 \hline
  & & & & &  \\ 
  \end{tabular}

\begin{tabular}{lccc} 
 $\psi(2S)$ $y$ bin &{ 2.0$-$3.0} & {3.0$-$3.5} & {3.5$-$4.5}  \\ 
 \hline 
 $W_+$ (GeV)& 772& 1115& 1634 \\ 
 $k_+ \mathrm{d}n/\mathrm{d}k_+(\times 10^{-3})$& 21.5& 18.5& 14.4 \\ 
 $r(W_+)$& 0.787& 0.762& 0.677 \\ 
 $W_-$ (GeV)& 63.4& 43.2& 29.9 \\ 
 $k_- \mathrm{d}n/\mathrm{d}k_-(\times 10^{-3})$& 45.3& 49.9& 52.4 \\ 
 $r(W_-)$& 0.911& 0.942& 0.926 \\ 
 $\sigma_{\gamma p\rightarrow \psi(2S) p(W_-)}$ (nb) & & &  \\ 
 Power law & 10.6& 8.2& 6.4 \\ 
 \hline 
 Calculated: & & &   \\ 
 $\sigma_{\gamma p\rightarrow \psi(2S) p(W_+)}$ (nb) & & &   \\ 
 Power Law & 64& 55& 88 \\ 
 \hline 
  \end{tabular}

\end{table}

Note that the correlation in the statistical covariance matrix is due to the conversion of the
statistical uncertainty on the reconstruced efficiency for each pseudorapidity
bin $\eta$ of the two muons to the rapidity bin $y$ of the \jpsi or $\psi(2S)$.

\begin{table}[!h]
 \caption{Gap survival factors for the production of \jpsi and \psitwos mesons at $\sqrt{s}=7$~TeV.}
 \label{tab:gap7}
\begin{center}
\begin{tabular}{lccccc}
\jpsi $y$ bin & {2.00$-$2.25} & {2.25$-$2.50} & {2.50$-$2.75}& {2.75$-$3.00} & {3.00$-$3.25} \\
\hline
$r(W_+) $ & 0.766 & 0.752 & 0.736 & 0.718 & 0.698 \\
$r(W_-) $ & 0.882 & 0.885 & 0.888 & 0.891 & 0.894 \\
& & & & &  \\
\jpsi $y$ bin & {3.25$-$3.50} & {3.50$-$3.75} & {3.75$-$4.00} & {4.00$-$4.25} & {4.25$-$4.50} \\
\hline
$r(W_+) $ & 0.676 & 0.650 & 0.620 & 0.587 & 0.550\\
$r(W_-) $ & 0.897 & 0.899 & 0.902 & 0.904 & 0.906 \\
& & & & &  \\
& & & & &  \\ 
\psitwos $y$ bin & {2.00$-$2.25} & {2.25$-$2.50} & {2.50$-$2.75}& {2.75$-$3.00} & {3.00$-$3.25} \\
\hline
$r(W_+) $ & 0.757 & 0.741 & 0.724 & 0.705 & 0.683 \\
$r(W_-) $  & 0.879 & 0.882 & 0.886 & 0.889 &0.892 \\
& & & & &  \\ 
\psitwos $y$ bin & {3.25$-$3.50} & {3.50$-$3.75} & {3.75$-$4.00} & {4.00$-$4.25} & {4.25$-$4.50} \\
\hline
$r(W_+) $ & 0.658 & 0.630 & 0.598 & 0.562 & 0.522 \\
$r(W_-) $  & 0.895 & 0.898 & 0.900 & 0.903 & 0.905 \\
\end{tabular}
\end{center}

\end{table}


\end{document}